\documentclass{article} 
\usepackage{graphicx}
\usepackage{subcaption}
\usepackage{nips14submit_e,times,scicite}
\usepackage{url}
\usepackage[english]{babel}
\usepackage[T1]{fontenc}
\usepackage[utf8]{inputenc}
\usepackage{listings}
\usepackage{float}
\usepackage{subfiles}
\providecommand{\keywords}[1]{\textbf{\textit{Keywords---}} #1}
\usepackage{booktabs} 
\usepackage{multirow}
\usepackage[export]{adjustbox}[2011/08/13]

\title{Creativity in temporal social networks: How divergent thinking is impacted by one's choice of peers}

\author{
Raiyan Abdul Baten\textsuperscript{1}, Daryl Bagley\textsuperscript{2}, Ashely Tenesaca\textsuperscript{2}, Famous Clark\textsuperscript{2},\\ \textbf{James P. Bagrow\textsuperscript{3}, Gourab Ghoshal\textsuperscript{4}, Ehsan Hoque\textsuperscript{2,*}}\\
\textsuperscript{1}Department of Electrical and Computer Engineering, University of Rochester, NY, USA\\ \textsuperscript{2}Department of Computer Science, University of Rochester, NY, USA\\ \textsuperscript{3}Department of Mathematics \& Statistics, University of Vermont, VT, USA\\
\textsuperscript{4}Department of Physics and Astronomy, University of Rochester, NY, USA\\
\texttt{*mehoque@cs.rochester.edu}
}

\nipsfinalcopy 

\begin{document}

\maketitle
\begin{abstract}
Creativity is viewed as one of the most important skills in the context of future-of-work. In this paper, we explore how the dynamic (self-organizing) nature of social networks impacts the fostering of creative ideas. We run $6$ trials ($N=288$) of a web-based experiment involving divergent ideation tasks. We find that network connections gradually adapt to individual creative performances, as the participants predominantly seek to follow high-performing peers for creative inspirations. We unearth both opportunities and bottlenecks afforded by such self-organization. While exposure to high-performing peers is associated with better creative performances of the followers, we see a counter-effect that choosing to follow the same peers introduces semantic similarities in the followers' ideas. We formulate an agent-based simulation model to capture these intuitions in a tractable manner, and experiment with corner cases of various simulation parameters to assess the generality of the findings. Our findings may help design large-scale interventions to improve the creative aptitude of people interacting in a social network.
\end{abstract}

\keywords{Network science, creativity, dynamic social networks}

\section*{}
Recent advances in robotics, AI and machine learning are increasingly focused on mimicking or even surpassing human capabilities. These innovations, however, have serious implications on our future workforce~\cite{frank2019toward}. Approximately $51\%$ of the tasks done in the US economy can be automated~\cite{manyika2017future}, and for each robot on the factory floor, some six jobs are lost~\cite{rotman2018making}. The need for manual labor in predictable and repetitive work is declining, while the demand is soaring for expertise in creative tasks, problem-solving, and other social-cognitive avenues of soft-skills~\cite{manyika2017future,alabdulkareem2018unpacking,baten2019upskilling}. Again, many of the critical and challenging tasks of the human civilization require humans to collaborate with others~\cite{wu2019large}, where they need to perform creatively at both individual and collective levels. Thus, enhancing the creative abilities of collaborating humans has become one of the aspirational challenges today. This motivation for creativity-at-scale leads to the exploration of social networks of creative collaborators. For instance, the development of an innovative product such as an aircraft or a computer operating system is only made possible by an interacting network of creative problem solvers, who benefit from each other's expertise~\cite{KAZANJIAN2000273}. Discussions in an academic network of researchers, faculty members and students can stimulate ideas for novel explorations. A graphic designer can find creative inspiration from peer-interactions in online networks like Reddit, Behance or Twitter. Adopting a social network lens helps us better understand the mechanisms, bottlenecks and opportunities for maximizing creative outcomes at-scale.

Researchers have examined the effects of various network attributes on creativity~\cite{perry2003social}. For instance, relationship strength, position and external ties of people are known to influence creative performance~\cite{perry2006social}. However, a key element missing from most prior literature is the dynamic characteristic of real social networks. Human interactions are structured in social networks, where people have control over who they interact with. Given an objective, they can choose to make or break ties to update the connectivity patterns around them~\cite{perc2010coevolutionary}, often in response to the behavior, performance, prestige, age, gender, popularity, self-similarity and other cues of the social partners~\cite{henrich2015secret,herrmann2007humans,boyd2011cultural}. This dynamic characteristic affords opportunities in human populations that static networks cannot: for example, dynamic networks promote cooperation~\cite{rand2011dynamic,szolnoki2008making}, collective intelligence~\cite{almaatouq2020adaptive}, and speaking skills~\cite{shafipour2018buildup}, among others. 

When it comes to creativity, the dynamic nature of social networks has received rather little research attention. Perry-Smith \textit{et al.} proposed a spiraling model, capturing the cyclical relationship between creativity and network position, where one fosters the other~\cite{perry2003social}. The argument being, if someone is creative at something, it might draw more attention to the creative person, resulting in an increased centrality and visibility. Again, a central person can inspire creative thoughts in others and also get inspired by others more readily than a peripheral person, thus helping in further creative ideation. However, this chain of argument has not been directly tested. Despite some efforts in examining such temporal effects from network and evolutionary game theoretic perspectives~\cite{jiang2018dynamic,armano2017beneficial,javarone2017solving}, it remains largely unclear how dynamic creative networks evolve with time, what laws they follow, and what implications such evolving has on creative ideation. 

This motivates the desire to understand how creative performances are exhibited and impacted in dynamic social networks. Consequently, in this paper, we first explore how connectivity patterns adapt to individual performance cues in a creativity-centric dynamic network. Second, we test how such adaptations impact creative ideation performances in a dynamic network, against the controls of static (fixed connections) and solo (unconnected) networks.

We run six trials of a web-based experiment, where participants in a dynamic social network performed idea generation tasks for $5$ rounds. The participants chose after each round which of the peers' ideas they wanted to be shown as stimuli (see Experimental setup). Following~\cite{perry2003social}, we anticipated that people looking for creative inspiration from others will use success cues to determine who among the peers are more creative and, therefore, more promising to be advantageous to form ties with (``follow''). We find that in the dynamic networks, the participants were indeed drawn towards following the most creative ideators. The statistical rarity and novelty ratings of one's ideas appear to be robust predictors of his/her popularity in the networks. 

If people preferentially form ties with the highly creative ideators in the network, what implications will it have on the creative outcomes? The associative theory of creative cognition suggests that priming people properly, e.g., by exposing them to others' ideas~\cite{siangliulue2015toward,chan2016comparing}, can stimulate their long-term memory circuitry. This can enable retrieving remotely stored concepts from memory~\cite{nijstad2003cognitive}. Combining such remote concepts can help in synthesizing novel ideas~\cite{paulus2007toward,brown2002making}. Based on this, we anticipated that following highly creative peers can enable better creative performance in people. For instance, following a person who generates rare ideas will increase the chances that the follower comes across ideas that have little overlap with his/her own. This can stimulate further ideas through novel association of concepts, resulting in ideas that would not have occurred to the follower otherwise~\cite{paulus2000groups,dennis2003electronic}. However, we also anticipated a counter-effect that if many people follow the same highly creative ideators in a dynamic network, the followers' inspiration sets will become overlapping, which might introduce redundancy in the stimulated ideas~\cite{burt2004structural,zhou2009social}. Our results show that following highly creative ideators is indeed associated with one's better creative performance. However, participants who chose to follow the same people (same stimuli) show an increasing semantic similarity in their independently stimulated ideas with time. These results suggest that self-organizing in a dynamic network can lead to conflicting opportunities and constraints. We formulate a simulation model that captures these empirically-derived intuitions, and helps assess the generality of the processes and insights.

\paragraph*{Experimental setup.} It is challenging to identify a dataset in the wild that offers traceable links between ideas and their stimuli in a temporal manner. We therefore resort to an artificial social network created in the virtual laboratory.

We are interested in divergent creativity, which signifies a person's ability to come up with numerous and varied responses to a given prompt/situation~\cite{kozbelt2010theories}. We use a customized version of Guilford's Alternate Uses Test~\cite{guildford1978alternate}, a widely-adopted approach for quantifying divergent creativity. In each of $5$ rounds, the participants were instructed to consider an everyday object (e.g., a brick), whose common use was stated (``used for building''). The participants needed to submit alternative uses for the object which are different than the given use, yet are appropriate and feasible. We choose the first $5$ objects from the Form B of Guilford's test as the ideation objects respectively in the $5$ rounds. 

\begin{figure}
    \centering
    \includegraphics[width=1\linewidth]{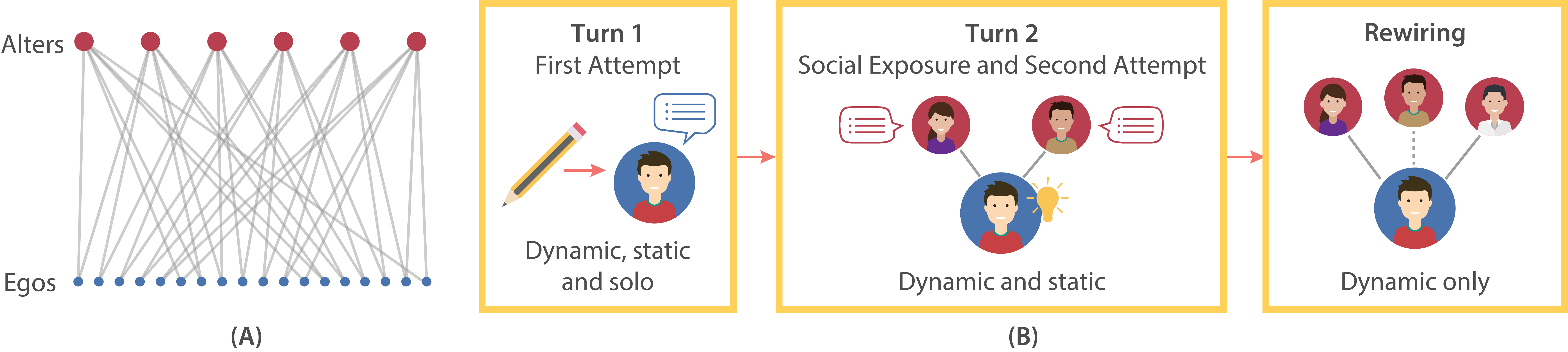}
    \caption[]{(A) The bipartite network structure used in the static condition, and as the initial configuration in the dynamic condition. (B) Study protocol for each round. In turn-$1$, the participants generated ideas independently. In turn-$2$, the dynamic and static egos received social exposure and could list further ideas. Only the dynamic egos could update which alters to follow at the end of each round.}
    \label{protocol_comb}
\end{figure}

We recruited $288$ participants from Amazon Mechanical Turk, who live in USA and are diverse in their age, racial, ethnic and gender distributions (see SI). We placed them randomly in one of three network conditions: (1) Dynamic, (2) Static, and (3) Solo. The static and solo conditions act as controls against which we assess the performances in the dynamic condition. For the dynamic and static conditions, we adopted a bipartite network structure (Figure~\ref{protocol_comb}A). There are two types of nodes in the network, \textit{alters} and \textit{egos}. First, we pre-recorded the ideas of $6$ \textit{alters}, who generated ideas independently. Then, we used these ideas as the stimuli for $36$ \textit{egos}---$18$ of them placed in a dynamic network condition, and the other $18$ in static. This bipartite design helped ensure a uniform stimuli for all the egos in the static and dynamic conditions. We repeated the process for $6$ independent trials, each with its unique alters. Under the solo condition, $36$ participants generated ideas in isolation.

Initially, the dynamic and static egos were connected to $2$ alters each using the network structure in Figure~\ref{protocol_comb}A. In each round, the egos first generated ideas independently for $3$ minutes (`turn 1'). They were then shown the ideas of the $2$ alters they were connected to, and given an additional $3$ minutes to list further ideas (`turn 2'). The egos were instructed not to resubmit any of the alters' exact ideas, and that only non-redundant ideas would contribute to their performance. They were also told that there will be a short test at the end of the study, where they will need to recall the ideas shown to them. This was to ensure that the participants paid attention to the stimuli, which is known to positively impact ideation~\cite{nijstad2006group}. Then, the egos rated the ideas of all of the $6$ alters on novelty (5-point ratings, 1: not novel, 5: highly novel). Finally, in the dynamic condition, the egos could optionally follow/unfollow alters at the end of each round to have an updated list of $2$ followees each. Except for the alters' username and ideas, no other information about the alters was shown to the egos. The egos were required to submit the rationale behind their choices of updating/not updating links in each round. This was in place to make the dynamic egos accountable for their choices, which is known to raise epistemic motivation and improve systematic information processing~\cite{bechtoldt2010motivated}. The egos in the static condition could not update their links, marking the only difference between the static and dynamic conditions. 

The participants in the solo condition were given $6$ minutes to list their ideas without any external stimuli. Detailed descriptions and examples guided the participants throughout the study. Everyone was paid $\$10$, and the top $5$ egos/solo participants (in each group of $18$) with the most number of non-redundant ideas were awarded $\$5$ bonuses. Figure~\ref{protocol_comb}B summarizes the protocol. 

\textbf{Measures.} We quantify creativity using three metrics: (1) Non-redundant idea counts, (2) Average novelty ratings and (3) Creativity quotient (see Methods for details). If an idea is given by at most a threshold number of participants in a given participant pool, it is considered non-redundant~\cite{oppezzo2014give}. The non-redundant idea count is thus a measure of the statistical rarity of the ideas.

Each idea in the dataset was rated on novelty by multiple people. We take the mean rating received by an idea as its novelty indicator, and consider the average novelty rating of a given set of ideas as a creativity metric.

The creativity quotient metric, $Q$, uses information theoretic measures to capture the fluency (quantity of ideas) and flexibility (the ability to generate a wide variety of ideas) of a set of ideas~\cite{bossomaier2009semantic,snyder2004creativity,seco2004intrinsic}. In all three metrics, higher values indicate better creative performances.

\section*{Results}

\begin{figure}
    \centering
    \includegraphics[width=1\linewidth]{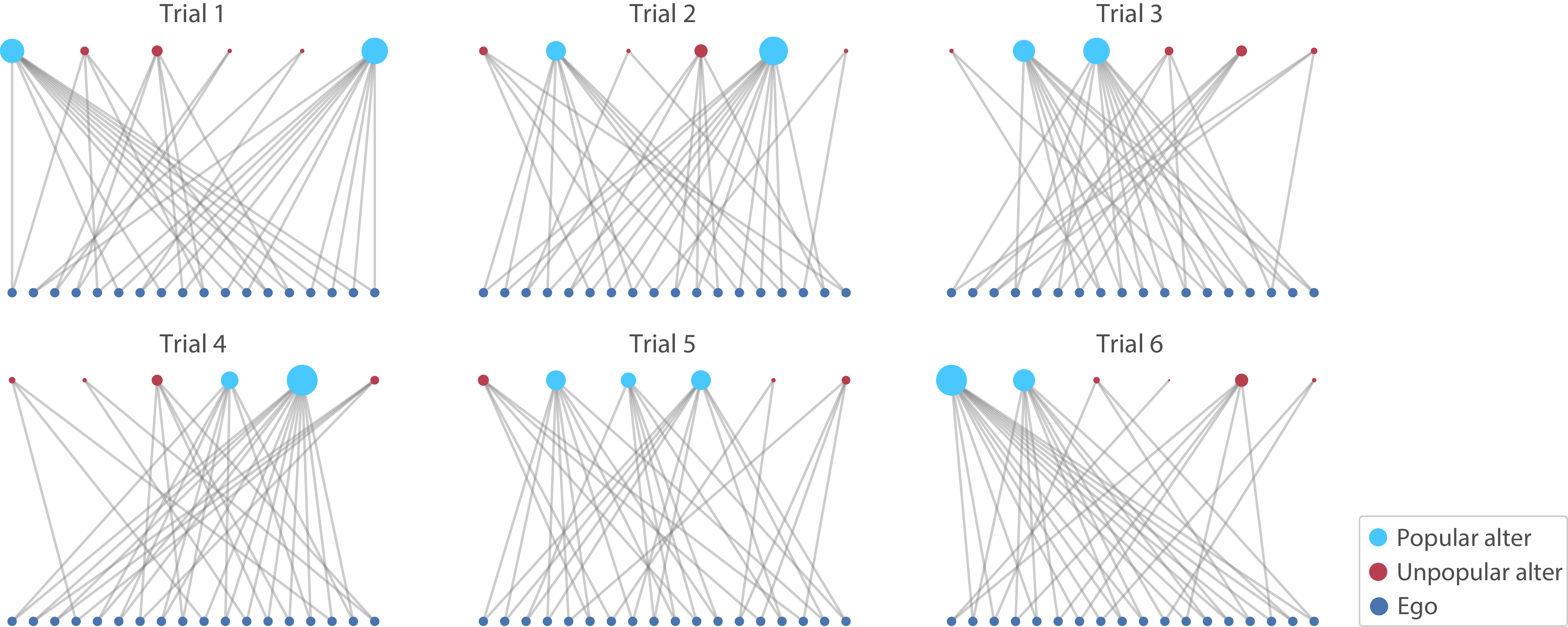}
    \caption[]{Final/evolved dynamic networks. The diameters of the circles in the upper-rows are proportional to the alters' follower counts at the end of the $5$\textsuperscript{th} round.}
    \label{network_evol}
\end{figure}

\begin{figure}
    \centering
    \includegraphics[width=0.7\linewidth]{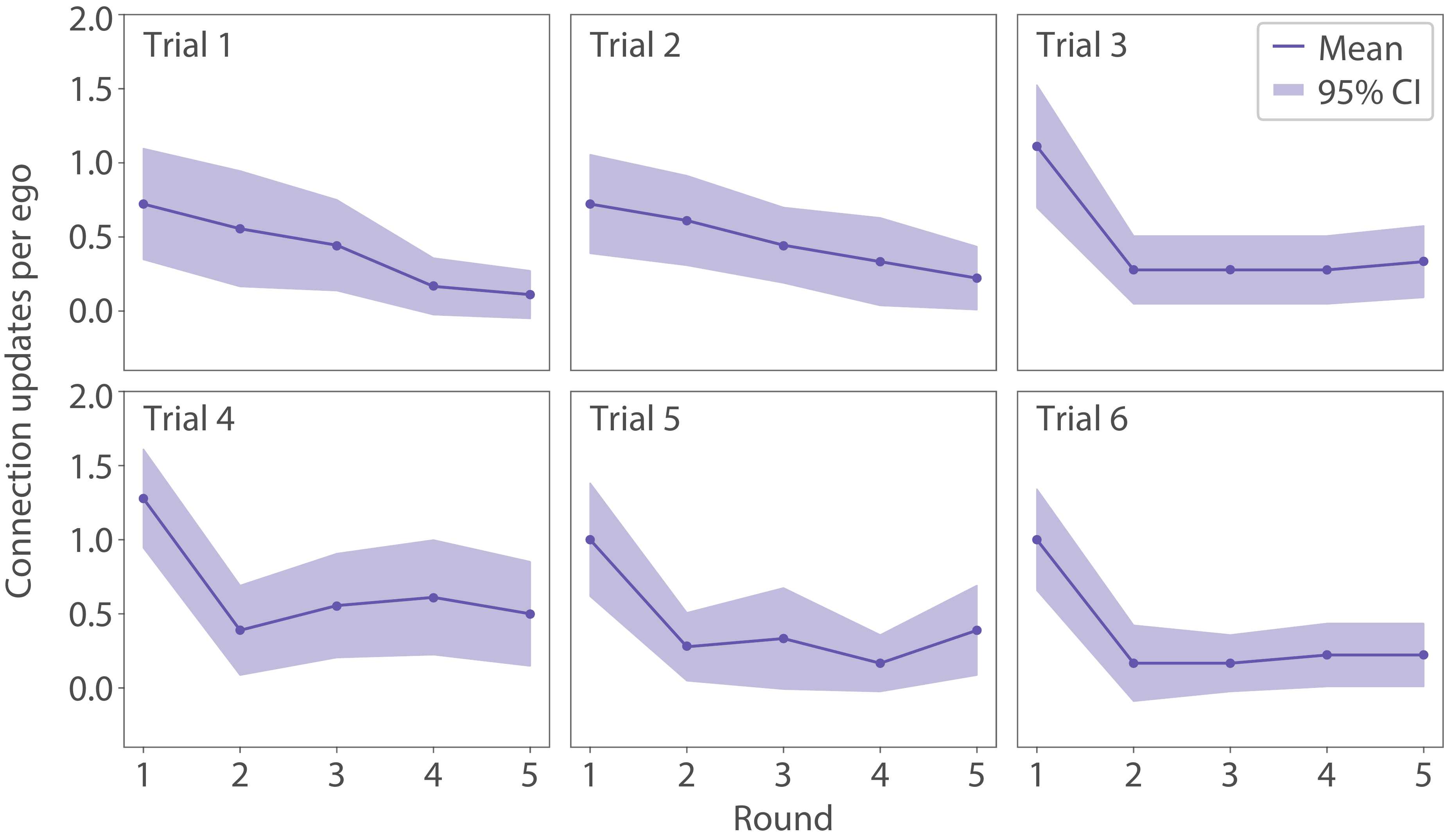}
    \caption[]{Number of connections updated per ego at the end of each round.}
    \label{conn_updates}
\end{figure}

\textbf{Link update patterns in the network evolution.} We first explore how connectivity patterns adapt to individual performance cues in creativity-centric dynamic networks. The networks evolved as the egos updated their lists of $2$ alters across the rounds. All of the alters started with $6$ ego-followers, but after $5$ rounds of network evolution, some of the alters lost followers, and some of them gained. The final evolved networks of the $6$ trials are shown in Figure~\ref{network_evol}. The number of connection updates per ego at the end of each round had a downward trend ($p<0.001$ for the negative slope, Figure~\ref{conn_updates}). Out of a maximum of $2$ possible updates, an average of $0.97$ connections were updated per ego after the first round, while $0.3$ connections were updated per ego after the fifth round. This suggests that as the egos received information about the alters' performances, they made up their minds on whom to follow, and readjusted later if necessary.

\begin{figure}
    \centering
    \includegraphics[width=1\linewidth]{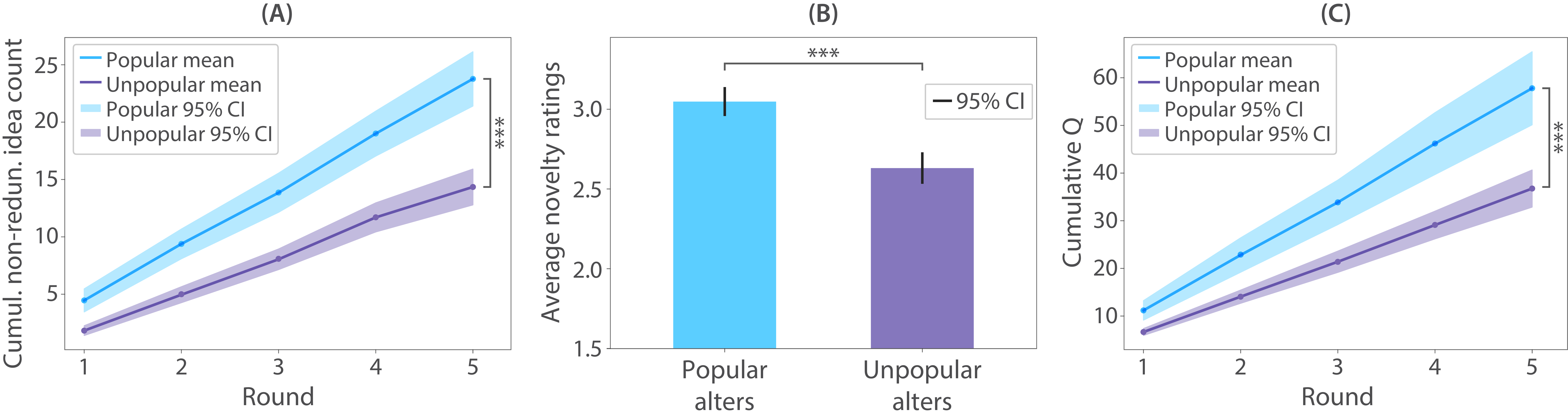}
    \caption[]{The popular alters significantly outperformed the unpopular alters in all three metrics: (A) The total number of non-redundant ideas in all $5$ rounds ($2$-tailed test, $t(34)=7.291$, $p<0.001$), (B) Average novelty ratings in all $5$ rounds ($t(34)=5.7$, $p<0.001$), and (C) The total creativity quotient in all $5$ rounds ($t(34)=5.81$, $p<0.001$). ***$p<0.001$.}
    \label{boss_lost_comb}
\end{figure}

We denote alters who finished with $>6$ and $\le6$ followers as `popular' and `unpopular' alters respectively. If the egos didn't update their links at all, the alters would still have the initially assigned $6$ followers, so we take $>6$ followers as a marker of popularity. The total non-redundant idea counts of the popular alters in all $5$ rounds were significantly higher than those of unpopular alters (popular (p) vs unpopular (u) alters: $2$-tailed test, $n_p=13$, $n_u=23$, $t(34)=7.291$, $p<0.001$, Figure~\ref{boss_lost_comb}A; $p<0.05$ in all $6$ trials, SI Figure S1). Similarly, the popular alters outperformed the unpopular alters significantly in the average novelty ratings ($2$-tailed test, $t(34)=5.7$, $p<0.001$, Figure~\ref{boss_lost_comb}B; $p<0.05$ in $5$ out of $6$ trials, SI Figure S2). The total creativity quotient $Q$ of the popular alters in all $5$ rounds were again significantly higher than those of unpopular alters ($2$-tailed test, $t(34)=5.81$, $p<0.001$, Figure~\ref{boss_lost_comb}C; $p<0.05$ in $3$ out of $6$ trials, SI Figure S3). SI Table S1 provides further details.

Using multivariate linear regression, we explore how the alters' creative performances correspond to their final popularity. The rational being, by the end of the fifth round, the egos had full knowledge of the qualities of the alters' ideas. So, we can expect the egos' final choices to be captured by the overall performances of the alters. As the dependent variable $y_i$, we take the fraction of egos connected to an alter $i$ at the end of the $5$\textsuperscript{th} round. The independent variables are: (1) the relative number of non-redundant ideas, $u'_i = \frac{u_i}{\sum_i u_i}$, (2) relative average novelty ratings, $\bar{r}'_i= \frac{\bar{r}_i}{\sum_i \bar{r}_i}$, and (3) relative creativity quotient, $Q'_i = \frac{Q_i}{\sum_i Q_i}$. Here, $u_i$, $\bar{r}_i$ and $Q_i$ are the total number of non-redundant ideas, average novelty ratings and total creativity quotients of alter $i$ in all $5$ rounds. We take the relative performance of the alters with respect to other alters in a given trial, since the egos could only choose from a fixed pool of alters. While all the independent variables correlate strongly with the dependent variable (Pearson's $\rho=0.80,0.86,0.75$ respectively, $p<0.001$ in each), multivariate regression helps us explore the relative contributions of the three independent variables. Mathematically,
\begin{equation}\label{future_ratings}
    y_i = \beta_0 + \beta_1 u'_i  + \beta_2 \bar{r}'_i + \beta_3 Q'_i   
\end{equation}

We train on $70\%$ data, compute $R^2$ and adjusted-$R^2$ on the remaining $30\%$ data, and repeat the process $10$ times (each time with a new 70/30 split) to compute a confidence interval around the mean $R^2$ and mean adjusted-$R^2$ values. Table~\ref{tab3} summarizes the results. We first test the independent variables separately using univariate regression, and find $\bar{r}'$ to give the best adjusted-$R^2_{\bar{r}'}=0.69$ (Model $1$). Adding $u'$ makes the adjusted-$R^2_{\bar{r}',u'}=0.72$ (Model $2$), while all three features give an adjusted-$R^2_{\bar{r}',u',Q'}=0.70$ (Model $3$). In Model $3$, the coefficients are significant for $\bar{r}'$ and $u'$, but not for $Q'$. This shows that the associations reported here are systematic for the first two predictors ($\bar{r}'$ and $u'$), which together explain $72\%$ of the variation in the dependent variable. Thus, the egos' connectivity dynamics are strongly captured by the novelty of the alters' ideas ($\bar{r}'$) and moderately by the statistical rarity of those ideas ($u'$). The relative creativity quotient metric has much of its information overlapped with the other two metrics. 

\begin{table}
\caption[]{Regression results of predicting the alters' relative popularity from their relative creativity markers. $\beta=$ standardized regression coefficient. **$p<0.01$. ***$p<0.001$. $N=36$ for each model.}
\begin{center}
    \begin{tabular}{|c|c c|c c|c c|}
    \hline
  \multirow{2}{*}{\textbf{Predictor}} &  \multicolumn{2}{|c|}{\textbf{Model 1: $\bar{r}'$ only}} &  \multicolumn{2}{|c|}{\textbf{Model 2: $\bar{r}'$, $u'$}} &  \multicolumn{2}{|c|}{\textbf{Model 3: all predictors}} \\
  \cline{2-7} 
   & $\beta$ & $t$ (std. err.) & $\beta$ & $t$ (std. err.)& $\beta$ & $t$ (std. err.)\\
    \hline
    $\bar{r}'$ & $0.1851$*** & $9.762$ $(0.019)$ & $0.1278$*** & $4.972$ $(0.026)$ & $0.1026$** & $3.435$ $(0.030)$ \\
    $u'$ & --- & --- & $0.0767$** & $2.983$ $(0.026)$ & $0.0704$** & $2.763$ $(0.025)$  \\
    $Q'$ & --- & --- & --- & --- & $0.0399$ & $1.562$ $(0.026)$  \\
    \hline
    $R^2$ & \multicolumn{2}{|c|}{$0.72$, $95\%$ C.I. =$[0.69,0.75]$} &  \multicolumn{2}{|c|}{$0.78$, $95\%$ C.I. =$[0.74,0.81]$} &  \multicolumn{2}{|c|}{$0.79$, $95\%$ C.I. =$[0.75,0.83]$}\\
    \hline
    Adjusted-$R^2$ & \multicolumn{2}{|c|}{$0.69$, $95\%$ C.I. =$[0.66,0.72]$} &  \multicolumn{2}{|c|}{$0.72$, $95\%$ C.I. =$[0.67,0.77]$} &  \multicolumn{2}{|c|}{$0.70$, $95\%$ C.I. =$[0.65,0.75]$}\\
     \hline
    \end{tabular}
    \label{tab3}
\end{center}
\end{table}

As most of the connection updates happened after the first round, we explore how the alters' first round performances can predict their popularity at the end of the first and fifth rounds. We find that the features $\bar{r}$ and $Q$ from the first round together give the best adjusted-$R^2=0.67$ in predicting the alters' popularity after the first round ($95\%$ C.I.=$[0.59,0.76]$). Similarly, for predicting the popularity after the fifth round only from the first round's performances, the feature $\bar{r}$ alone gives the best adjusted-$R^2=0.60$ ($95\%$ C.I.=$[0.52,0.67]$). This drop in the explainability of the dependent variable can be due to the fact that when we try to predict the final popularity of the alters from the first round's data, we miss out on the performance information from the middle rounds, which can have an effect on the ego's final choices. When we incorporate all $5$ rounds' data, we are able to predict the final popularities with adjusted-$R^2$ as high as $0.72$ (Model $2$).

The key take-away is that the egos in a dynamic network predominantly form ties with the consistently high-performing alters, as anticipated. In typical creativity studies, participants are shown stimuli randomly or based on intrinsic qualities~\cite{chan2016comparing,siangliulue2015toward}. Here, a key contrast is that the networks are allowed to dynamically self-organize, as one can choose for oneself who to take inspirations from. Thus, the implications of such adaptations on the ideation process, as explored below, become direct manifestations of the dynamic nature of the networks.

\textbf{Exposure to high-performing alters is associated with better creative performances of the egos.} We argued previously that forming ties with high-performing alters should increase the chances that an ego comes across ideas that have little overlap with his/her own. This lack of overlap, in turn, can increase the chances of stimulating new ideas by facilitating novel associations between remote concepts. To test this, we take advantage of the fact that in turn-1, the egos generated ideas independently prior to any social exposure, which allows us to test the overlap their ideas have with their alters' ideas (measured by Jaccard index, see Methods). In turn-2, the egos could see the stimuli, allowing us to explore whether the creative qualities of the stimulated ideas have any association with how creative the respective alters were. 

We first find the round-wise popular alters by identifying those with $>6$ followers at the end of each round. Then, for each round, we split the egos into three `groups', where (i) both, (ii) only one, and (iii) none of the followees of an ego are round-wise popular. Thus, we analyze each ego's data across $5$ rounds, where the ego belongs to one of these $3$ `groups' in each round. The group sample sizes are $n_i= 273$, $n_{ii}= 476$, and $n_{iii}= 331$ respectively. We assess the fixed effects of the group and round factors ($3$ and $5$ levels respectively) against the random effects of the egos to control for repeated measures. We employ the Aligned Rank Transform (ART) procedure, which is a linear mixed model-based non-parametric test~\cite{wobbrock2011aligned}, available in the ARTool package in R~\cite{artool}. 

 \begin{figure}[t]
    \centering
    \includegraphics[width=1\linewidth]{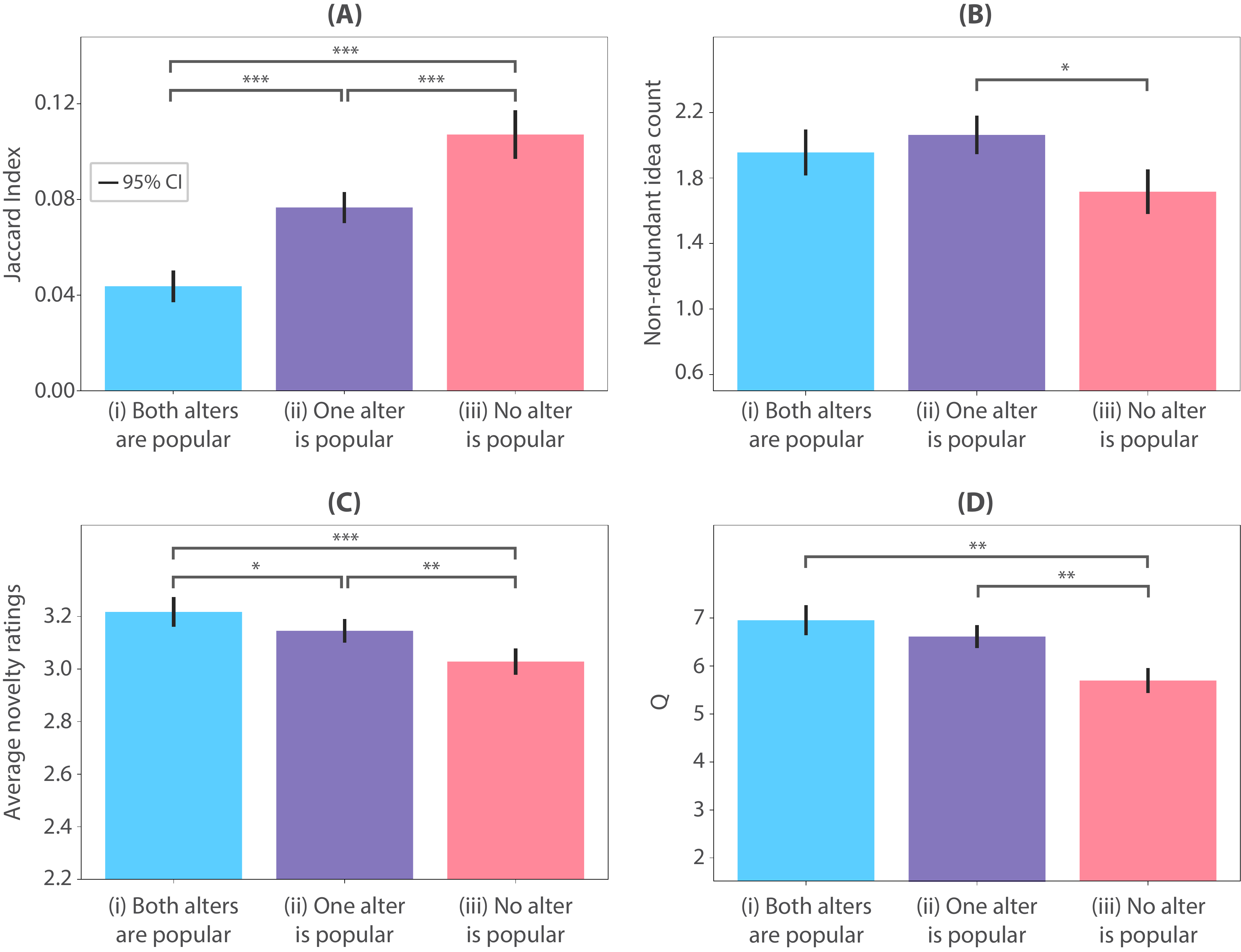}
    \caption[]{(A) Average overlap between the egos' turn-1 ideas and their alters' ideas. We compare among egos with (i) both, (ii) only one, and (iii) no round-wise popular alters. Following more high-performing alters decreases the egos' exposure to overlapping ideas. Panels (B), (C) and (D) show the creative performances of the same groups of egos in turn-2. In all three metrics, egos in group (iii) perform the worst. *$p<0.05$, **$p<0.01$, ***$p<0.001$; all $p$-values corrected for multiple comparisons and repeated measures.}
    \label{echo_comb1}
\end{figure}

We analyze the average overlap between each ego's turn-1 ideas and his/her two alters' ideas. We find a significant main effect of the group factor (ART procedure, $F(2,669.84)=66.53$, $p<0.001$). Post-hoc pairwise comparisons among the group factor levels in the fitted ART model show that group (i) had significantly less idea-overlap than group (ii) ($t(887)=-7.52$, $p<0.001$), and group (ii) had significantly less overlap than group (iii) ($t(574)=-5.56$, $p<0.001$). All $p$-values are Bonferroni-corrected. Thus, following more high-performing alters systematically decreased the overlap between an ego's turn-1 ideas and the alters' ideas, as anticipated (Figure~\ref{echo_comb1}A and SI Tables S2, S3). Group (i) consistently had less overlap than groups (ii) and (iii) in each round (SI Figure S4).

We then explore the creativity measures of the turn-2 ideas across the three levels of the group factor. We fit three separate models for the three metrics. All of the three creativity metrics show significant main effects for both the group and round factors, but no significant interaction between the factors (Main effects of the group factor: (1) Non-redundant idea counts, $F(2,825.86)=3.7$, $p=0.025$, (2) Average novelty ratings, $F(2,535.47)=11.85$, $p<0.001$, (3) Creativity quotient, $F(2,1036.36)=6.66$, $p=0.001$). 

We proceed to conduct post-hoc analysis among the $3$ group levels as before. We find that group (iii), i.e., egos who followed unpopular alters only, showed the worst performance in all three metrics. In particular, group (ii) significantly outperformed group (iii) in non-redundant idea counts ($t(727)=2.69$, $p=0.02$), but the other two pair-wise comparisons (group (i) vs group (ii) and group (i) vs group (iii)) were insignificant in this metric. In case of the average novelty ratings, group (i) significantly outperformed group (ii) ($t(721)=1.97$, $p=0.049$), and group (ii) significantly outperformed group (iii) ($t(458)=3.4$, $p=0.002$). As for creativity quotient, both groups (i) and (ii) outperformed group (iii) (respectively, $t(1022)=3.06$, $p=0.005$ and $t(1016)=3.5$, $p=0.002$). There was no significant difference between groups (i) and (ii). See Figure~\ref{echo_comb1}B-\ref{echo_comb1}D and SI Tables S4-S5. These results imply that following at least one high-performing alter is associated with better creative performance of the egos. Our results thus suggest evidence for better stimulation of ideas when egos are exposed to high-quality ideas.

\textbf{Following the same alters introduces semantic similarities in the egos' ideas.} We further motivated a counter-argument that if multiple egos follow the same alters, their stimuli sets will become overlapping. This can make the egos' stimulated ideas similar, despite ideating independently. To test this, we explore whether the semantic (dis)similarities between each node-pair's turn-2 ideas have any association with the number of common alters they have. We estimate the semantic nature of ideas using neural word embeddings (Word2Vec~\cite{mikolov2013efficient}) and compare the dissimilarity of the embeddings using Word Mover's Distance~\cite{kusner2015word} (see Methods).

\begin{figure}
    \centering
    \includegraphics[width=1\linewidth]{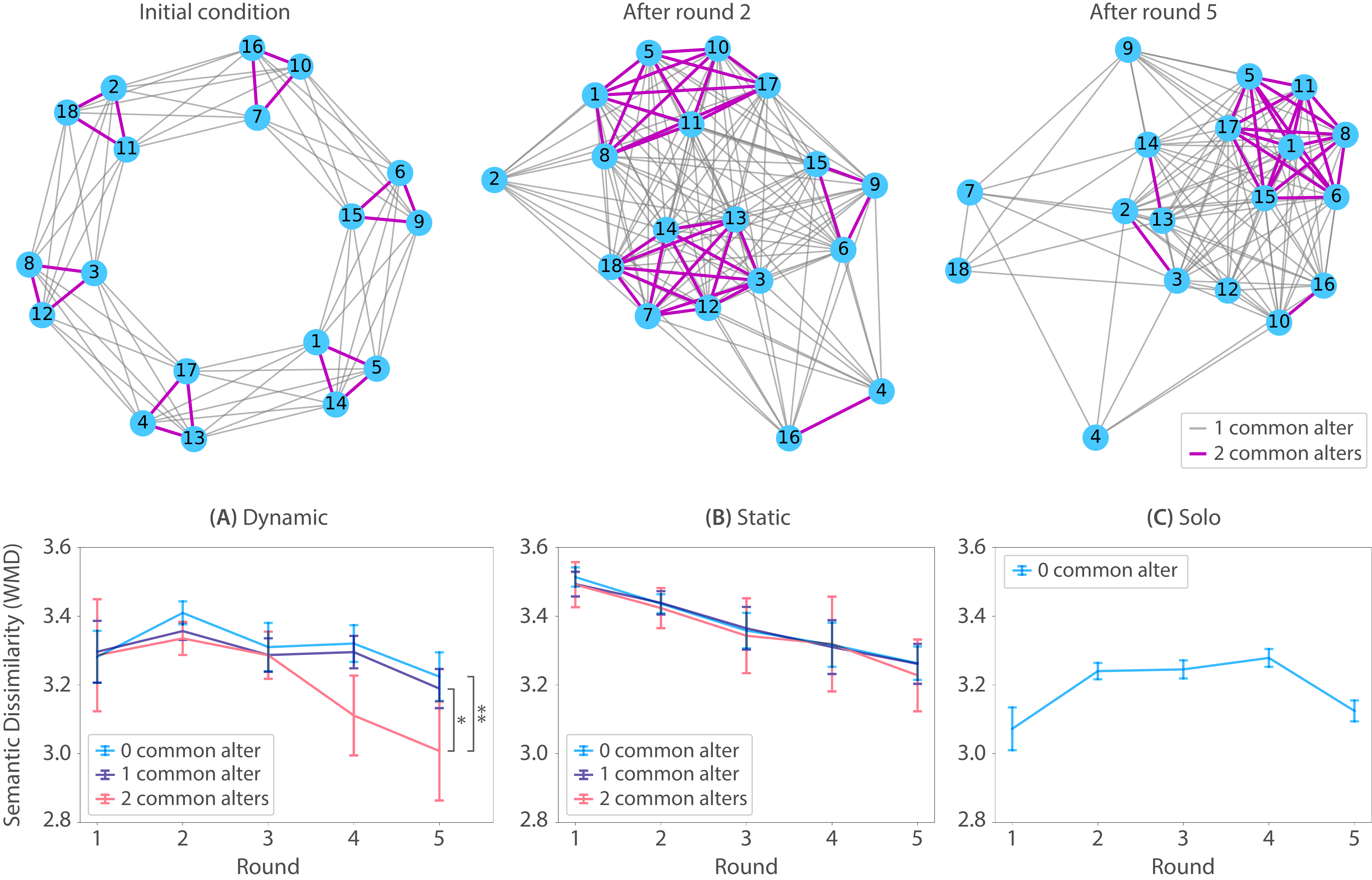}
    \caption[]{\textit{(Top row)} Example one-mode projections of the bipartite networks on the ego-nodes from trial 4. In the static condition, the initial condition remained fixed. \textit{(Bottom row)} Semantic dissimilarity (Word Mover's Distance, WMD) between the idea-sets of node-pairs are shown for (A) dynamic, (B) static and (C) solo conditions. Node-pairs that share two common alters in the dynamic condition show significantly less dissimilarity by the fifth round than the $0$ and $1$ common alter cases (2-tailed tests, $2$ vs $0$ common alter(s): $t(452)=-2.962$, $p<0.01$; $2$ vs $1$ common alter(s): $t(632)=-2.788$, $p<0.02$). Whiskers denote $95\%$ C.I. *$p<0.05$, **$p<0.01$ (Bonferroni-corrected).}
    \label{echo_comb}
\end{figure}

We first take one-mode projections of the round-wise bipartite networks on the ego nodes. In the projected graphs, two ego-nodes are connected with an edge if they have common alters (top row of Figure~\ref{echo_comb}). We compute the semantic dissimilarity between each node-pair's turn-2 idea-sets. We then compare the dissimilarities among node-pairs with $2$, $1$ and $0$ common alters (corresponding to the purple, gray and missing edges respectively in the projected graphs). 

As the dynamic-egos rewired their connections across rounds, the turn-2 ideas of node-pairs with $2$ common alters gradually became less dissimilar ($p<0.05$ for the negative slope, Figure~\ref{echo_comb}A). Node-pairs with $0$ and $1$ common alters did not show any such decreasing trend. At the end of the $5$\textsuperscript{th} round, the node-pairs with $2$ common alters were significantly less dissimilar than the $0$ and $1$ common alter cases (sample size of node-pairs: $n_2 = 170$, $n_1= 464$, $n_0= 284$; 2-tailed test; $2$ vs $0$ common alter(s): $t(452)=-2.962$, $p<0.01$; $2$ vs $1$ common alter(s): $t(632)=-2.788$, $p<0.02$; Bonferroni-corrected $p$-values; see SI Table S6). In the static condition, the alters were the same, but the network remained fixed. All of the three comparison cases of $0$, $1$ and $2$ common alter node-pairs showed steadily decreasing dissimilarity ($p<0.001$ for the negative slope in all three cases), but there was no difference among the three comparison cases ($p>0.05$, Figure~\ref{echo_comb}B). In the solo condition, there was no stimuli, and the semantic dissimilarity did not have any systematic trend ($p=0.68$ for the slope, Figure~\ref{echo_comb}C). 

This shows that as the rounds progressed, the ideas of egos who followed the exact same alters increasingly grew similar. Importantly, this effect is different from groupthink~\cite{nemeth2003better}, where the desire for harmony or conformity results in consensus among group members. In our case, the egos acted independently without the knowledge of other egos' ideas, yet became increasingly similar to the co-followers of the same alter.

\begin{figure}
    \centering
    \includegraphics[width=1\linewidth]{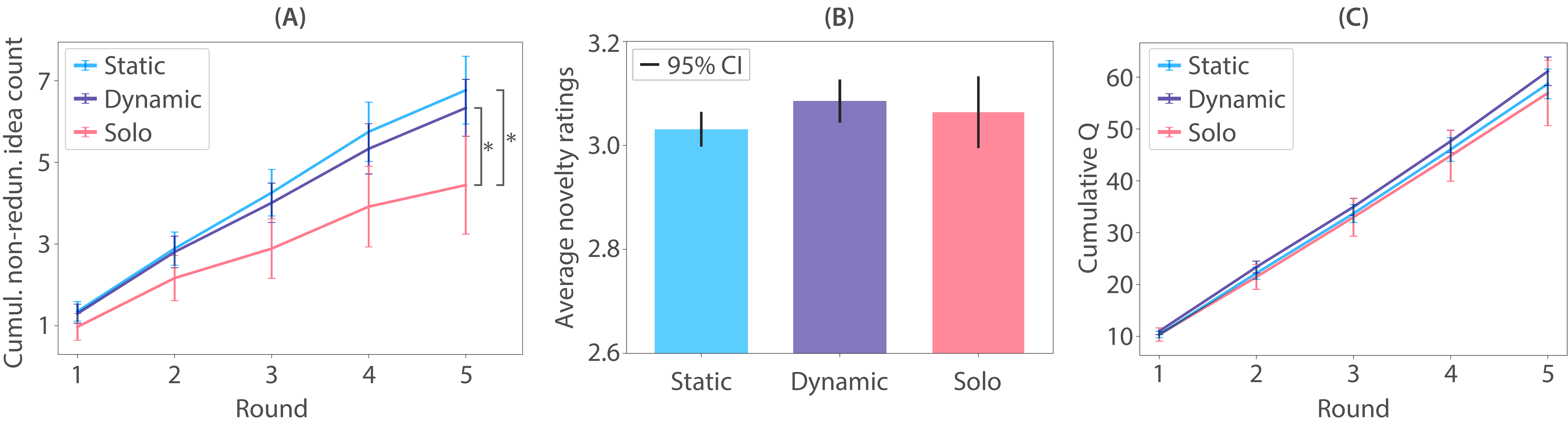}
    \caption[]{Individual-level comparisons of (A) cumulative non-redundant idea counts, (B) average novelty ratings and (C) cumulative creativity quotients among various study conditions. The static and dynamic conditions significantly outperformed the solo condition in the total number of non-redundant ideas ($2$-tailed test; dynamic vs solo: $t(142)=2.7$, $p<0.03$; static vs solo: $t(142)=2.898$, $p<0.02$). Whiskers denote $95\%$ C.I. *Bonferroni-corrected $p<0.05$.}
    \label{ind_comb}
\end{figure}

\textbf{Individual creative performance comparisons among various study conditions.} We analyze the individual creative performances in various study conditions. The participants in both the dynamic and static conditions significantly outperformed the solo participants in the total number of non-redundant ideas ($2$-tailed test; dynamic $n_d= 108$, static $n_s= 108$, solo $n_c= 36$; dynamic (d) vs solo (c): $t(142)=2.7$, $p<0.03$; static (s) vs solo (c): $t(142)=2.898$, $p<0.02$; all $p$-values Bonferroni-corrected; Figure~\ref{ind_comb}A, SI Table S7). However, there was no significant difference between the dynamic and static conditions ($2$-tailed test, $p>0.05$).

The dynamic egos showed significantly higher average novelty ratings than the static egos (2-tailed test, $p<0.05$), but after Bonferroni correction, the difference was no longer significant (Figure~\ref{ind_comb}B). The other condition-pair comparisons (solo vs dynamic and solo vs static) did not show any significant difference ($p>0.05$). The dynamic egos significantly outperformed the static egos in two trials out of six (2-tailed test, $p<0.05$).

The creativity quotient metric did not show any significant difference between any of the condition-pairs ($p>0.05$ for each condition pair, Figure~\ref{ind_comb}C). SI Figures S5-S7 summarize trialwise results.

Thus, at an individual level, we observe no consistent and systematic benefit in dynamic networks compared to their static counterparts. This is in contrast to another important human performance avenue---collective intelligence---where dynamic link adaptations have been shown to have performance benefits over the static condition~\cite{almaatouq2020adaptive}. However, in typical collective intelligence tasks, people can imitate their peers’ answers to get closer to the `correct' responses. In our study, the task encouraged \textit{open-ended} ideation, and none of the three creativity metrics captured any systematic benefit of the dynamic condition.

\textbf{Simulation model for the observed processes.} We formulate an agent-based simulation model for the observed processes. The purpose is two-fold: \textit{First,} to capture the network dynamics and the associated micro- and macro-level stimulation effects in a simple and tractable manner. This helps to provide clarity to the interplay between the network and cognitive processes in the system. \textit{Second,} to explore corner case outcomes of various simulation variables, e.g., the rewiring probability, inter-ego similarity, cognitive stimulation function and network size. These result in insights that enrich the empirical findings. The simulation model, analysis and results are elaborated in the SI. Below, we summarize the key ideas.

We first generate idea-sets for alter-agents such that some alters have more non-redundant ideas than others (capturing popular and unpopular alters). Starting from the same initial network structure as the empirical setup, the ego-agents gradually rewire their connections to increasingly follow the popular alters. We use a rewiring probability parameter $P_r$, where $P_r=0$ (no rewiring) results in the initial network structure, and $P_r=1$ (full rewiring) results in all of the egos following only the top performing alters. Thus, $P_r$ captures the network's temporal evolution.

We simulate the consequent network-driven and cognitive effects on the egos' ideation using three key processes: (A) As rewiring takes place, more egos connect to the popular alters. This is a \textit{network-driven} process, which makes the stimuli sets overlapping for egos who follow the same alters. (B) Given the stimuli set, the egos can generate stimulated ideas, which is a \textit{cognitive} process. Our empirical results show a positive association between the rarity of the stimuli and the generation of novel ideas. To capture this, we consider three abstractions of the cognitive process, where the statistical rarity of a stimulus idea affects the number of stimulated novel ideas in linear, sub-linear and super-linear manners. (C) There can be similarities in the stimulated ideas of independently ideating egos who share the same alters. This is again a \textit{network-driven} process, as the similarity is initiated/facilitated by the egos' similar choices of peers. We consider two extreme cases: no inter-ego similarity and full similarity, where all the ideas inspired by the same stimulus are mutually different and exactly the same, respectively. Importantly, if any of these three key processes is taken away, one cannot fully capture the insights from our empirical results.

\begin{figure}
    \centering
    \includegraphics[width=1\linewidth]{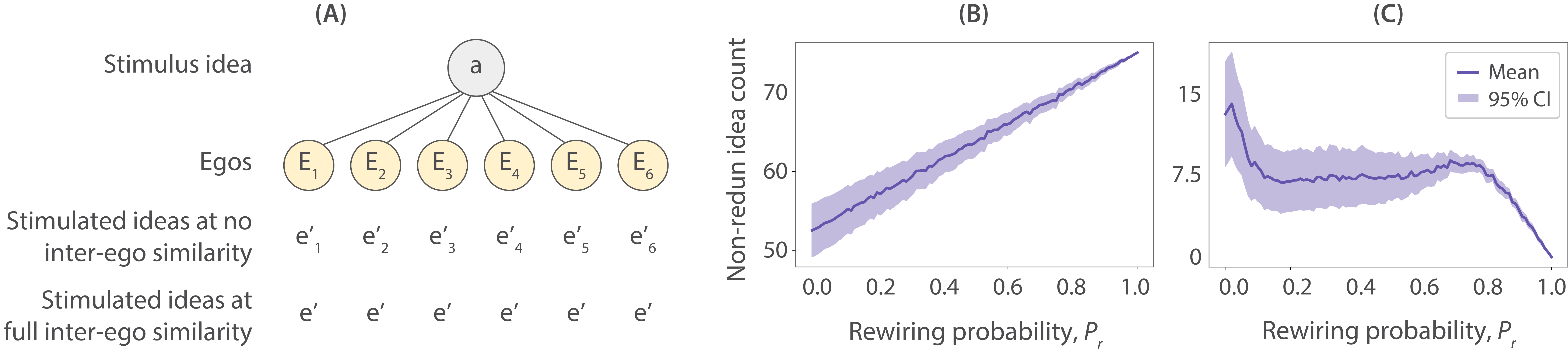}
    \caption[]{(A) Illustration of inter-ego similarity. Egos $E_1$ to $E_6$ view the same stimulus $a$, and independently generate a single idea each. At no inter-ego similarity, each of the stimulated ideas is mutually different. At full inter-ego similarity, all of the stimulated ideas become the same. Panels (B) and (C) show the total non-redundant idea counts in a network with $m=18$ alters and $n=54$ egos. As the rewiring probability $P_r$ increases, more egos follow the same highly popular alters in the network. In the absence of inter-ego similarity (panel B), an increase in $P_r$ leads to a better generation of non-redundant ideas. At full inter-ego similarity (panel C), $P_r=1$ makes all the stimulated ideas mutually redundant.}
    \label{simul_summary}
\end{figure}

Using the model, we explore various corner cases of network rewiring, inter-ego similarity, cognitive stimulation and network size variables. We find that if there is no inter-ego similarity whatsoever, process C loses relevance, and the cognitive stimulation mechanisms in process B become the key to the creative outcomes of the ego-agents. As $P_r$ increases, more novel ideas are then generated due to better stimulation, and the network's creative outcomes peak at $P_r=1$. On the other hand, at full inter-ego similarity, process B loses relevance, as all the stimulated ideas from a given stimulus become exactly the same. At $P_r=1$, all of the egos have the same stimuli. As a result, none of the stimulated ideas are rare anymore and the net creative output of the network drops to zero. While our empirical results of inter-ego similarity lie in between these extreme cases, the simulation nonetheless clarifies the opportunities and constraints in a systematic manner (Figure~\ref{simul_summary}).

These insights are robust to the various cognitive stimulation functions we considered in process B. We find that as long as the cognitive stimulation function captures a positive stimulation effect of a high quality/rare stimulus, the same outcomes are observed irrespective of whether the effect is linear, sub-linear or super-linear. We further experimented with various network sizes and found the effects to be robust in networks $3$, $10$ and $100$ times larger than our experimental ones. A detailed theoretical treatment of the model and the associated processes, for example beginning with a mean-field analysis, remain part of our future work.

\section*{Discussion}

Social cues and heuristics are ubiquitously used by humans for navigating through societal interactions, and contribute to their immense success as a species. Here, we first explored how the connectivity patterns in a creativity-centric dynamic network adapt to people's performance cues. From $6$ independent trials, we found that the egos' following/unfollowing patterns are governed significantly by the novelty (average novelty ratings) and statistical rarity (non-redundant idea counts) of the alters' ideas. These two metrics lead to an adjusted-$R^2=0.72$ in predicting the relative popularity of the alters. Perry-Smith \textit{et al.}'s spiral model suggests that highly creative people will enjoy increased visibility in a dynamic network~\cite{perry2003social}. Our results validate that idea and explain the relevant cues governing such tie formations. These dynamics are different from preferential attachment, since the egos were blind to the existing follower counts/popularity of the alters. The use of a bipartite network helped us keep the egos' stimuli-sets uniform and understand the link update patterns cleanly. However, pre-recording the alters' ideas made the network unidirectional, prohibiting us from testing the full spiral model. 

Although the alters were passive actors, our results have implications for them as well. Consider social media influencers, who act as third-party endorsers and shape audience perception through blogs, tweets, and other social media channels. The rise of such micro-celebrity has inspired a lot of research (e.g., \cite{khamis2017self}), and corporate brands are interested to leverage the marketing potential therein. We find that the alters need to generate not only statistically rare ideas, but also ideas of high quality and novelty to win more attention than others. This has implications for the influencers regarding how to stay relevant and ahead of others, in parallel to the Red Queen hypothesis~\cite{derfus2008red}.

We confirmed that following high-performing alters is associated with better creative performance of the egos. As the egos followed more high-performing alters, the overlap reduced between the ego's own ideas and the alters' ideas---which can partly explain the positive stimulation of ideas. In the dynamic networks, the egos showed a pattern of flocking behind the high-performing alters, thereby improving their own chances of generating novel ideas. However, there was a counter-effect that ego-pairs who followed the same alters in the dynamic condition had an increasing semantic similarity with time. At the end of the fifth round, they had significantly higher semantic similarity than ego-pairs who did not have both of their alters in common. This shows one way network dynamicity can hurt original idea generation: choosing the same stimuli may inadvertently and increasingly make people's ideas similar to each other, despite independent ideation. Importantly, these processes are driven by the egos' own choices of alters, capturing the dynamic nature of real social networks. Our simulation model captures these interplay between the network-driven and cognitive processes.

These insights can lead to research efforts towards making social networks more creatively competent. Consider academicians who follow the same popular domain-experts on social media, seeking inspirations for novel research. Indeed, high quality stimuli can help the followers generate high quality ideas themselves. But at the same time, there can be similarities in the stimulated ideas of the followers. Social network intervention strategies will then need to strike a sweet spot to help the followers get the best out of their networks, e.g., by recommending diversified sets of followees to people. If the high preforming alters act as gravitational force attracting and maintaining the egos’ connections, the semantic similarities among the followers' contents could potentially signal when the network needs intervention for re-stimulating ideation. The same arguments can apply to offline social networks as well, e.g., in large-scale creative teams. This can be particularly pertinent for groups known to gravitate towards strong personalities. As for an ego, there are implications that seeking out high-quality inspirations can be worthwhile, but flocking behind popular people might not always be optimal for one to stand out.

Drawing from relevant literature, our study settings were designed to reduce performance bottlenecks and increase cognitive stimulation of the egos. The key bottlenecks known to affect performance in brainstorming sessions~\cite{paulus2000groups,dennis2003electronic} were not present here: there was no evaluation apprehension from peers (alters) due to asynchronous exposure, no social loafing as the individuals were responsible for their own performance, and no production blocking as the egos could think on their own in turn-1. Furthermore, the quiz at the end and the justifications for connection update decisions recorded each round were in place to increase cognitive stimulation and and epistemic motivation~\cite{nijstad2006group,paulus2000groups,bechtoldt2010motivated}. As expected, the dynamic and static conditions, which had stimuli from alters, enjoyed a significantly higher count of non-redundant ideas compared to the solo condition, which had no stimulus.

To make the findings robust, we repeated our experiment for $6$ independent trials. We reported robust out-of-sample $R^2$ and adjusted-$R^2$ results. The simulation model further helped us assess the generality of the insights under various corner cases. However, our study is not without limitations. The steep expenses associated with conducting the experiment prohibited us from collecting a larger dataset. We conducted power analyses to ensure that the reported results have sufficient statistical power (see SI). The participants generated alternative uses of common objects, which hardly resembles real-life creative challenges. The study settings prohibited us from exploring the effects of bidirectional creative influence. Also, the study lasted for $5$ rounds, which can be prohibitively short to demonstrate the full effects of network dynamicity. Longitudinal studies with practical creative challenges can generate further insights on the research questions. 

\section*{Methods}\label{s_methods}
\subsection*{Creativity Metrics}
We operationalized creativity using the three following metrics, based on previous literature.
\subsubsection*{Non-redundant idea counts} To compute the non-redundant idea counts~\cite{oppezzo2014give}, we first rejected submissions that were infeasible and not different from the given use. Then, the round-wise idea-sets were organized/binned to collect the same ideas together. For binning, we followed the coding rules specified in~\cite{bouchard1970size} and in the scoring key of Guilford's Alternate Uses test\footnote{Copyright @ 1960 by Sheridan Supply Co., all rights reserved, published by Mind Garden, Inc, www.mindgarden.com.}. 

Then, we computed the statistical rarity of the ideas. An idea was determined to be non-redundant if it was given by at most a threshold number of participants in a given pool of ideas. For the alters, the threshold was set to $1$, and the pool was set to be the round-wise idea-set of the $6$ alters in the trial. When comparing trial-wise dynamic and static egos, the threshold was heuristically set to $2$, with the pool being the round-wise idea-set of the $36$ egos. In other words, if $\ge3$ egos in a trial submitted the same idea, it was no longer considered non-redundant. When comparing the data of the solo, static and dynamic conditions aggregated over all trials, the threshold was once again heuristically set to $2$, and the pool was set to all the ideas generated by all these participants. 

$2$ research assistants independently binned similar ideas together from the first $3$ trials. They were shown randomly ordered anonymized ideas. The non-redundant idea counts of the participants computed based on their coding had a high agreement (Krippendorff's $\alpha = 0.85$; Spearman's $\rho=0.92$, $p<0.001$, $95\%$ C.I.$=[0.885, 0.941]$). Then, the first research assistant coded the rest of the dataset, which was used in the analyses.

\subsubsection*{Creativity Quotient}

Creativity Quotient, $Q$, accounts for both the quantity of ideas generated and the quantity of distinct categories those ideas fall into~\cite{snyder2004creativity,bossomaier2009semantic}. If the ideas of a participant are very similar, they are likely subtle variations of a small number of categories. Conversely, if they are very dissimilar, they likely touched many categories---marking better creativity. 

The computation of $Q$ uses an information theoretic measure of semantic similarity derived from WordNet~\cite{miller1995wordnet}. Concepts appear as syn-sets (synonym sets) in WordNet, and the nouns come with an `is a' relationship. We first remove stopwords and punctuations from the ideas, and run a spell-checker. We then split each idea into its constituting set of concepts, and replace verbs and adjectives with related nouns whenever possible. Then, we find the information content of each of those concepts. Since the taxonomic structure of WordNet is organized in a meaningful way, concepts with many hyponyms should convey less information than the ones with a small number of hyponyms~\cite{seco2004intrinsic}. Thus, infrequent concepts (e.g., leaf nodes) should hold more information than the abstracting nodes. We therefore quantify the Information Content, $I$, of a concept $c$ as,
\begin{equation}
    I(c) = \frac{log\big(\frac{h(c)+1}{w}\big)}{log\big(\frac{1}{w}\big)} = 1- \frac{log(h(c)+1)}{log(w)}
\end{equation}
where $h(c)$ is the number of hyponyms of concept $c$ and $w$ is the total number of concepts in the taxonomy. The denominator normalizes the metric with respect to the most informative concept, to have $I\in[0,1]$. 

We then determine how similar a given pool of ideas (from a participant in a given round) are. We compute the semantic similarities between every pair of concepts, $c_1$ and $c_2$, in the pool~\cite{jiang1997semantic} using 
\begin{equation}
    sim(c_1,c_2)= 1-\Big(\frac{I(c_1)+I(c_2)-2\times sim_{MSCA}(c_1,c_2)}{2} \Big)
\end{equation}
Here, the semantic similarity, $sim(c_1,c_2)$, is a function of the amount of information the two concepts have in common, $sim_{MSCA}(c_1,c_2)$. This, in turn, is given by the information content of the Most Specific Common Abstraction (MSCA) that subsumes both the concepts:
\begin{equation}
    sim_{MSCA}(c_1,c_2) = \max_{c'\in S(c_1,c_2)} I(c') 
\end{equation}
where $S(c_1,c_2)$ is the set of concepts subsuming $c_1$ and $c_2$.

We then compute the multi-information, $I_{m}$, as the shared information across the response-set. This is computed by first obtaining the maximum spanning tree from the network of concept similarity values between concept pairs, and then summing over the edge weights in the max spanning tree. Finally, $Q$ is obtained by,
\begin{equation}
    Q = N-I_m
\end{equation}
where $N$ is the total number of concepts in the idea pool.

\subsubsection*{Ratings} 

Each idea of the alters was rated on novelty by $36$ egos in the trial, generating $38,880$ ratings in total. The mean rating received by idea $j$ is taken as the idea-level rating $r_j$. For participant $i$ in round $p$, we compute the average novelty rating $\bar{r}_i^{(p)} = mean(r_j)$ for ideas $j$ by $i$ in $p$. We take the mean rating received by each alter from each ego, and compute the consequent intra-class correlation coefficient among the ego-raters in that trial. The mean intra-class correlation coefficient from all $6$ trials was $ICC(3,36)=0.945$.

Additionally, each idea of the egos and solo participants was rated by at least $4$ raters, hired separately from Amazon Mechanical Turk, resulting in $40,320$ ratings from $141$ raters. For fairness, each rater focused on one round in a given trial, and rated all the $36$ egos'/solo participants' ideas in that round. The raters were first given $3$ minutes to generate ideas on that round's prompt themselves, to familiarize them with the task. Then they rated randomly presented ideas, guided by instructions and examples. We find a positive mean intra-class correlation coefficient among the raters, $ICC(3,4)=0.317$.

\subsection*{Measure of idea overlap}
To measure the overlap between idea-sets $A$ and $B$, we use the Jaccard Index:
\begin{equation}
    J(A,B)=\frac{|A\cap B|}{|A\cup B|}.
\end{equation}
If $A=B=\emptyset$, we take $J(A,B)=1$. We use the binning from non-redundant idea count computation to identify same entries in the idea-sets.

\subsection*{Measure of semantic dissimilarity}
To measure the semantic dissimilarity of two idea-sets, we use the Word Mover's Distance~\cite{kusner2015word}. We first remove stop-words and punctuation. The dissimilarity of two idea-sets is computed by the minimum Euclidean distance that the Word2Vec~\cite{mikolov2013efficient} embedded words of one idea-set need to travel to reach the embedded words of another idea-set.

\section*{Author Contributions}
R.A.B. designed the study, collected, annotated, and analyzed the data, and authored the manuscript. D.B. analyzed the data. A.T. designed the study and annotated the data. F.C. designed the study and built the data collection platform. J.P.B. contributed to the data analysis and preparation of the manuscript. G.G. oversaw the study design, interpretation of results, and preparation of the manuscript. M.E.H oversaw the study design, interpretation of results, and preparation of the manuscript.

\section*{Acknowledgment}
We appreciate Shagun Bose, Ethan Otto and Xiaoning Guo's contributions to the project. 

\section*{Data Accessibility}
See \url{https://github.com/ROC-HCI/Creativity-in-temporal-social-networks} for data and code.  Due to copyright protection of the creativity test, we provide processed data of the participants' ideas rather than the raw data.

\section*{Ethics}
This study was approved by the Institutional Review Board at the University of Rochester, New York, USA.

\section*{Funding Statement}
This work was supported by NSF grants IIS-1750380 and IIS-1447634, and a Google Faculty Research Award.


\renewcommand{\thefigure}{S\arabic{figure}}
\renewcommand{\thetable}{S\arabic{table}}
\setcounter{figure}{0}
\setcounter{table}{0}

\clearpage
\setcounter{page}{1}
\date{}

\title{Supplementary Information\\ \textit{Creativity in temporal social networks: How divergent thinking is impacted by one's choice of peers}}
\newpage
\part*{Supplementary Information}

\maketitle
\textbf{Published in the Journal of the Royal Society Interface, October 2020}\\

\tableofcontents
\listoffigures
\listoftables
\newpage
\section{Demographic information}
Among the $288$ participants we recruited from Amazon Mechanical Turk, $167$ were male and $121$ were female. Their ages ranged from $18$ to $55$+ (18y-24y: $30$, 25y-34y: $129$, 35y-44y: $81$, 45y-54y: $23$, 55y+: $25$). The racial distribution was: White: $224$, Asian: $15$, Black or African American: $22$, American Indian or Alaska Native: $15$, other: $12$. Among them, $15$ participants belonged to Hispanic or Latino ethnicity.

\section{Supplementary figures}

\begin{figure}[H]
    \centering
    \includegraphics[width=1\linewidth]{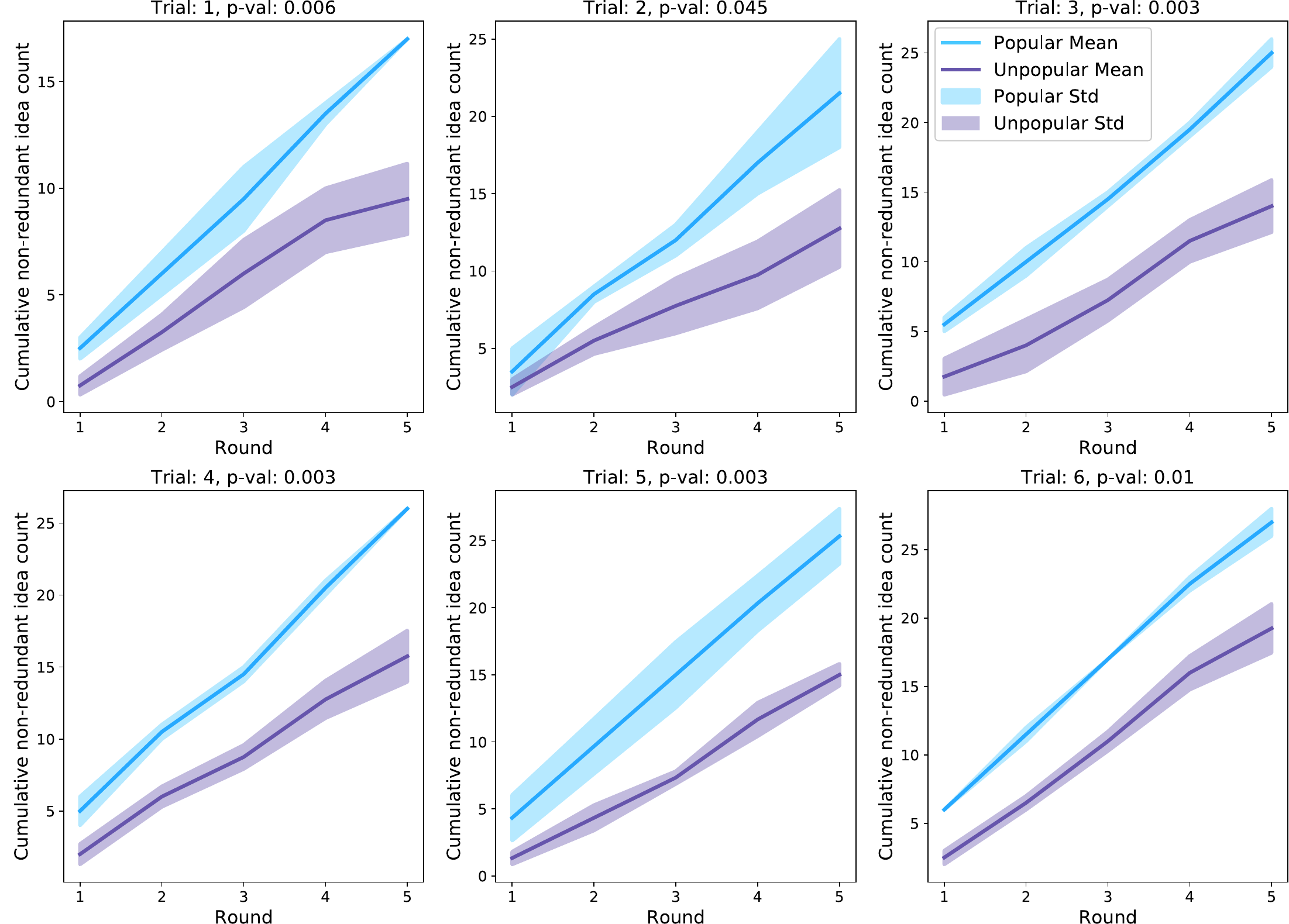}
    \caption[Trial-wise comparisons of cumulative non-redundant idea counts between popular and unpopular alters]{Trial-wise comparisons of cumulative non-redundant idea counts between popular and unpopular alters. 2-tailed tests show the popular alters (p) to have significantly higher cumulative counts over all rounds than unpopular alters (u) in all $6$ trials, detailed as follows. Trial 1: $m_p=17.0$, $m_u=9.5$, $t(4)=5.222$, $p=0.0064$, $95\%$ C.I. for $m_p-m_u=[4.0, 11.0]$; Trial 2: $m_p=21.5$, $m_u=12.8$, $t(4)=2.879$, $p=0.045$, $95\%$ C.I. for $m_p-m_u=[2.1, 15.4]$; Trial 3: $m_p=25.0$, $m_u=14.0$, $t(4)=6.351$, $p=0.0031$, $95\%$ C.I. for $m_p-m_u=[6.9, 15.1]$; Trial 4: $m_p=26.0$, $m_u=15.8$, $t(4)=6.629$, $p=0.0027$, $95\%$ C.I. for $m_p-m_u=[6.5, 14.0]$; Trial 5: $m_p=25.3$, $m_u=15.0$, $t(4)=6.609$, $p=0.0027$, $95\%$ C.I. for $m_p-m_u=[6.8, 13.9]$; Trial 6: $m_p=27.0$, $m_u=19.3$, $t(4)=4.66$, $p=0.0096$, $95\%$ C.I. for $m_p-m_u=[3.8, 11.7]$.}
    \label{SI1}
\end{figure}

\begin{figure}[H]
    \centering
    \includegraphics[width=1\linewidth]{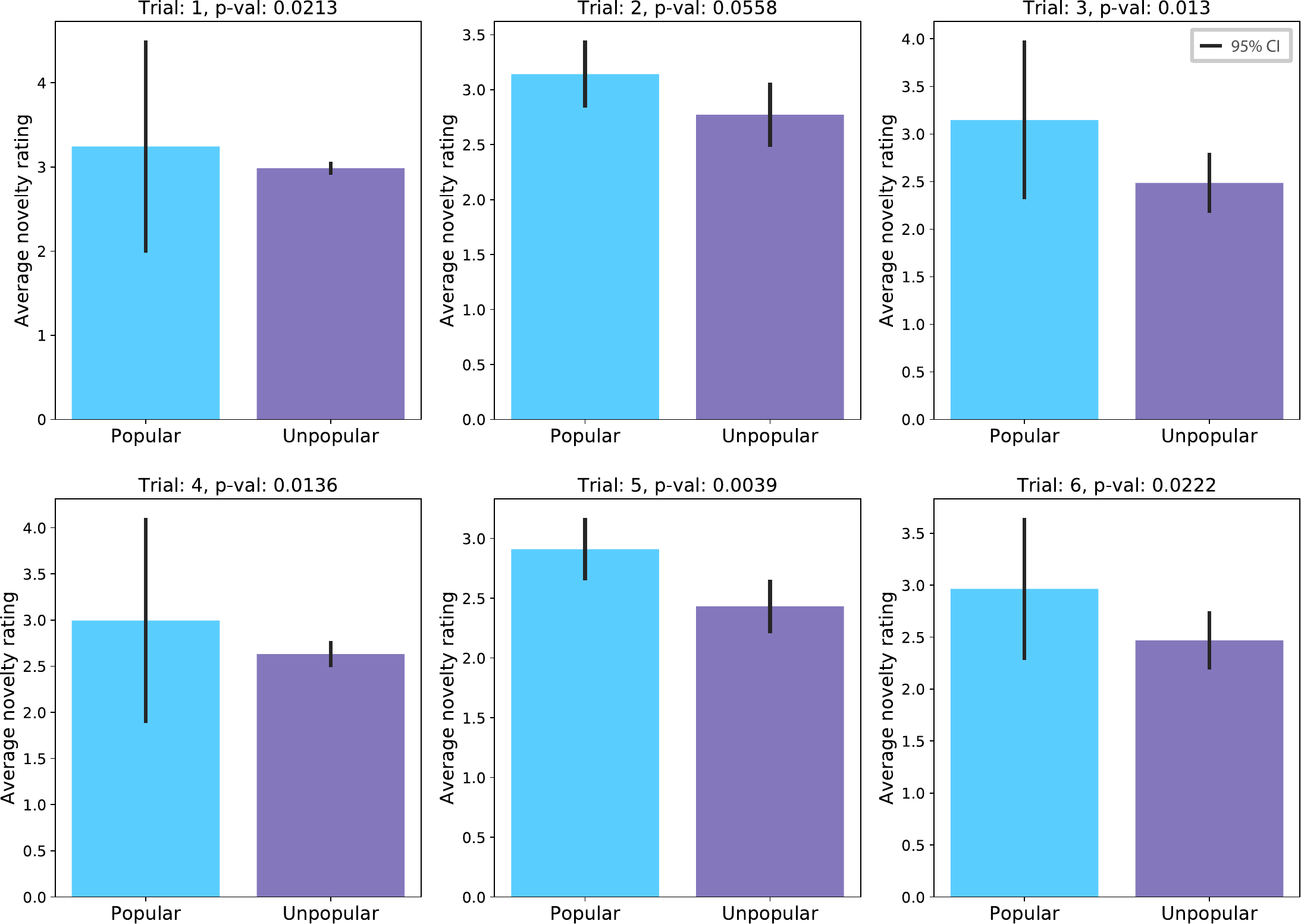}
    \caption[Trial-wise comparisons of average novelty ratings between popular and unpopular alters]{Trial-wise comparisons of average novelty ratings between popular and unpopular alters. 2-tailed tests show the popular alters (p) to have significantly higher average novelty ratings over all rounds than unpopular alters (u) in $5$ out of $6$ trials, detailed as follows. Trial 1: $m_p=3.2$, $m_u=3.0$, $t(4)=3.675$, $p=0.021$, $95\%$ C.I. for $m_p-m_u=[0.1, 0.4]$; Trial 2: $m_p=3.1$, $m_u=2.8$, $t(4)=2.67$, $p=0.0558$, $95\%$ C.I. for $m_p-m_u=[0.04, 0.7]$; Trial 3: $m_p=3.2$, $m_u=2.5$, $t(4)=4.264$, $p=0.013$, $95\%$ C.I. for $m_p-m_u=[0.3, 1.0]$; Trial 4: $m_p=3.0$, $m_u=2.6$, $t(4)=4.207$, $p=0.0136$, $95\%$ C.I. for $m_p-m_u=[0.2, 0.6]$; Trial 5: $m_p=2.9$, $m_u=2.4$, $t(4)=5.98$, $p=0.0039$, $95\%$ C.I. for $m_p-m_u=[0.3, 0.7]$, Trial 6: $m_p=3.0$, $m_u=2.5$, $t(4)=3.63$, $p=0.022$, $95\%$ C.I. for $m_p-m_u=[0.2, 0.8]$.}
    \label{SI2}
\end{figure}

\begin{figure}[H]
    \centering
    \includegraphics[width=1\linewidth]{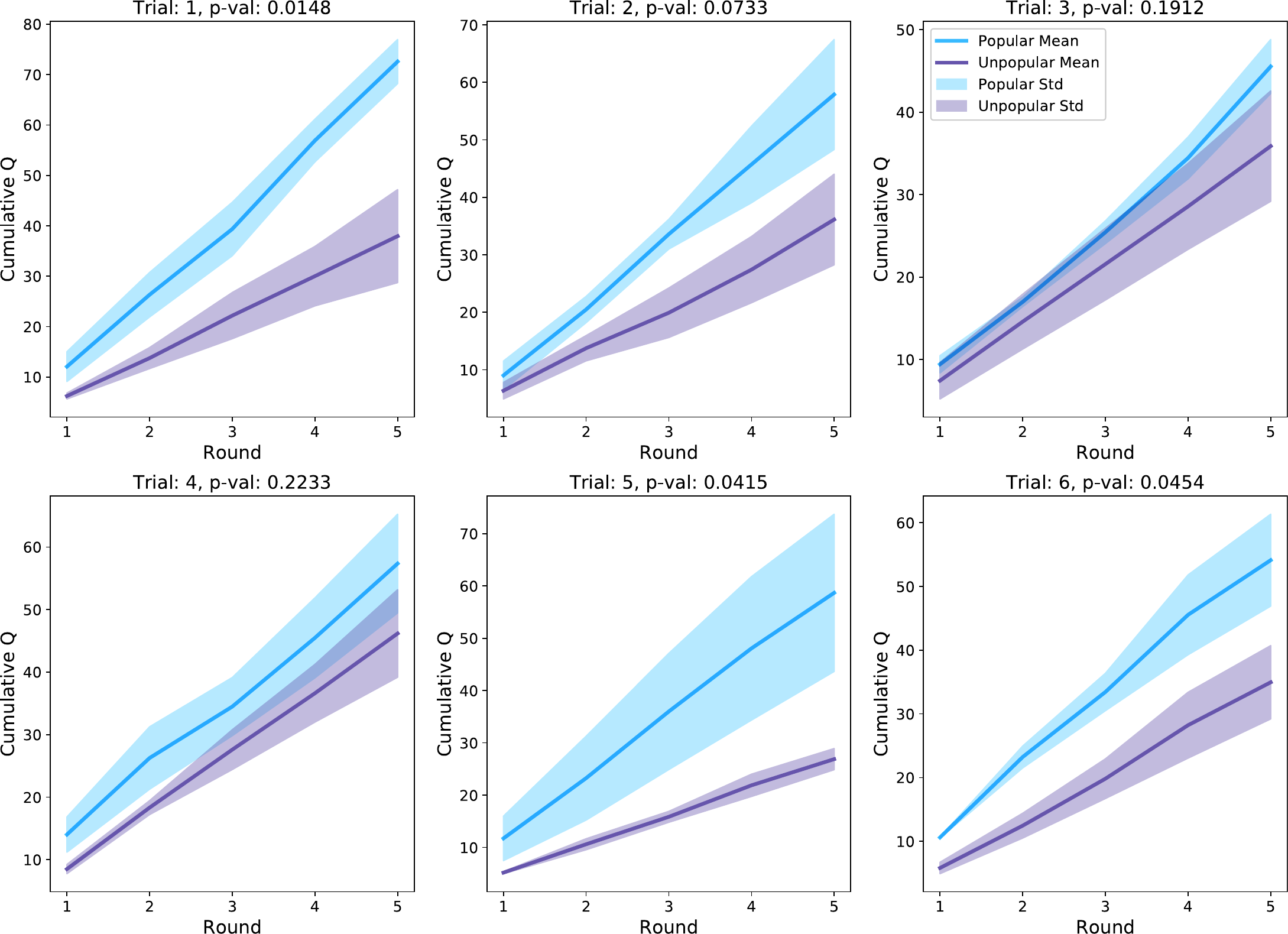}
    \caption[Trial-wise comparisons of cumulative $Q$ between popular and unpopular alters]{Trial-wise comparisons of cumulative $Q$ between popular and unpopular alters. 2-tailed tests show the popular alters (p) to have significantly higher total $Q$ scores over all rounds than unpopular alters (u) in $3$ of the trials, detailed as follows. Trial 1: $m_p=72.6$, $m_u=38$, $t(4)=4.102$, $p=0.015$, $95\%$ C.I. for $m_p-m_u=[14.7, 54.6]$; Trial 2: $m_p=57.9$, $m_u=36.1$, $t(4)=2.41$, $p=0.073$, $95\%$ C.I. for $m_p-m_u=[1.7, 41.8]$; Trial 3: $m_p=45.5$, $m_u=35.9$, $t(4)=1.572$, $p=0.19$, $95\%$ C.I. for $m_p-m_u=[-4.9, 24.2]$; Trial 4: $m_p=57.4$, $m_u=46.2$, $t(4)=1.44$, $p=0.223$, $95\%$ C.I. for $m_p-m_u=[-6.2, 28.6]$; Trial 5: $m_p=58.7$, $m_u=26.9$, $t(4)=2.962$, $p=0.041$, $95\%$ C.I. for $m_p-m_u=[7.5, 56.2]$; Trial 6: $m_p=54.1$, $m_u=35$, $t(4)=2.872$, $p=0.045$, $95\%$ C.I. for $m_p-m_u=[4.3, 34.0]$.}
    \label{SI3}
\end{figure}

\begin{figure}[H]
    \centering
    \includegraphics[width=0.6\linewidth]{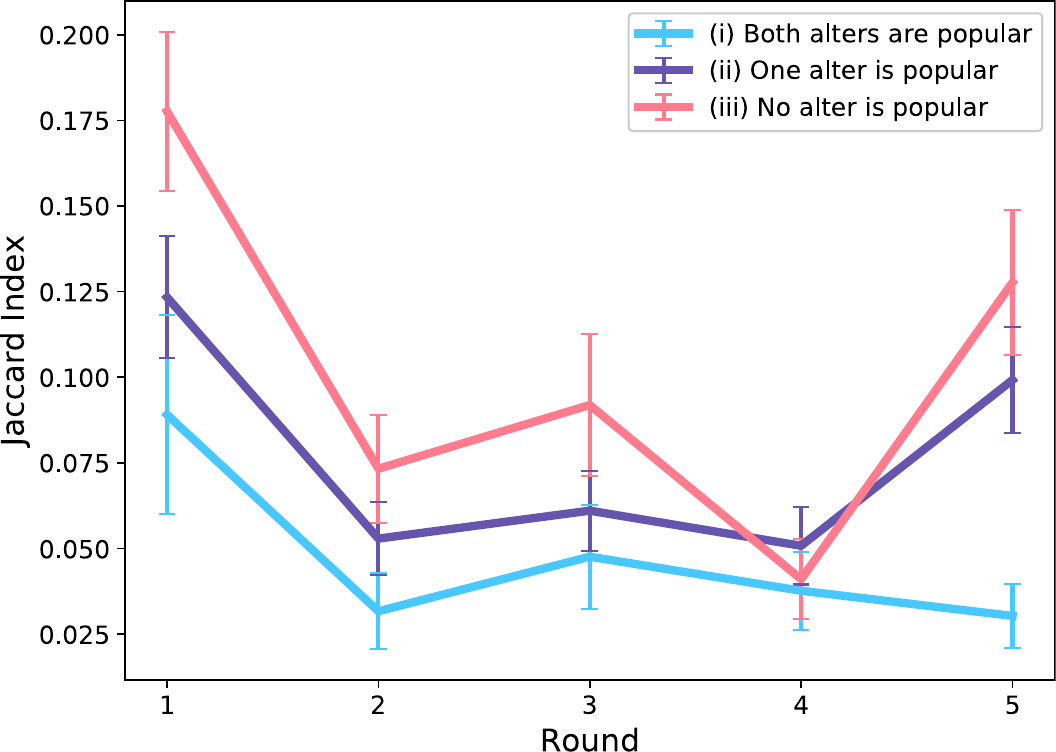}
    \caption[Average overlap between idea-sets of egos' turn-1 ideas and their alters in various rounds]{Average overlap (measured with Jaccard Index) between idea-sets of egos' turn-1 ideas and their alters in various rounds. Comparisons are made among three cases of egos: those with (i) both, (ii) only one and (iii) no alter(s) who are round-wise popular. As can be seen, egos who follow $2$ popular alters consistently show a lower overlap compared to the other two cases. 2-tailed test results on the fifth round is given below. (i) vs (ii): $m_1=0.03$, $m_2=0.1$, $t(145)=-7.03$, Bonferroni-corrected $p<0.001$, $95\%$ C.I. for $m_1-m_2=[-0.088, -0.05]$; (i) vs (iii): $m_1=0.03$, $m_3=0.13$, $t(131)=-8.223$, Bonferroni-corrected $p<0.001$, $95\%$ C.I. for $m_1-m_3=[-0.121, -0.074]$.}
    \label{SI4}
\end{figure}

\begin{figure}[H]
    \centering
    \includegraphics[width=1\linewidth]{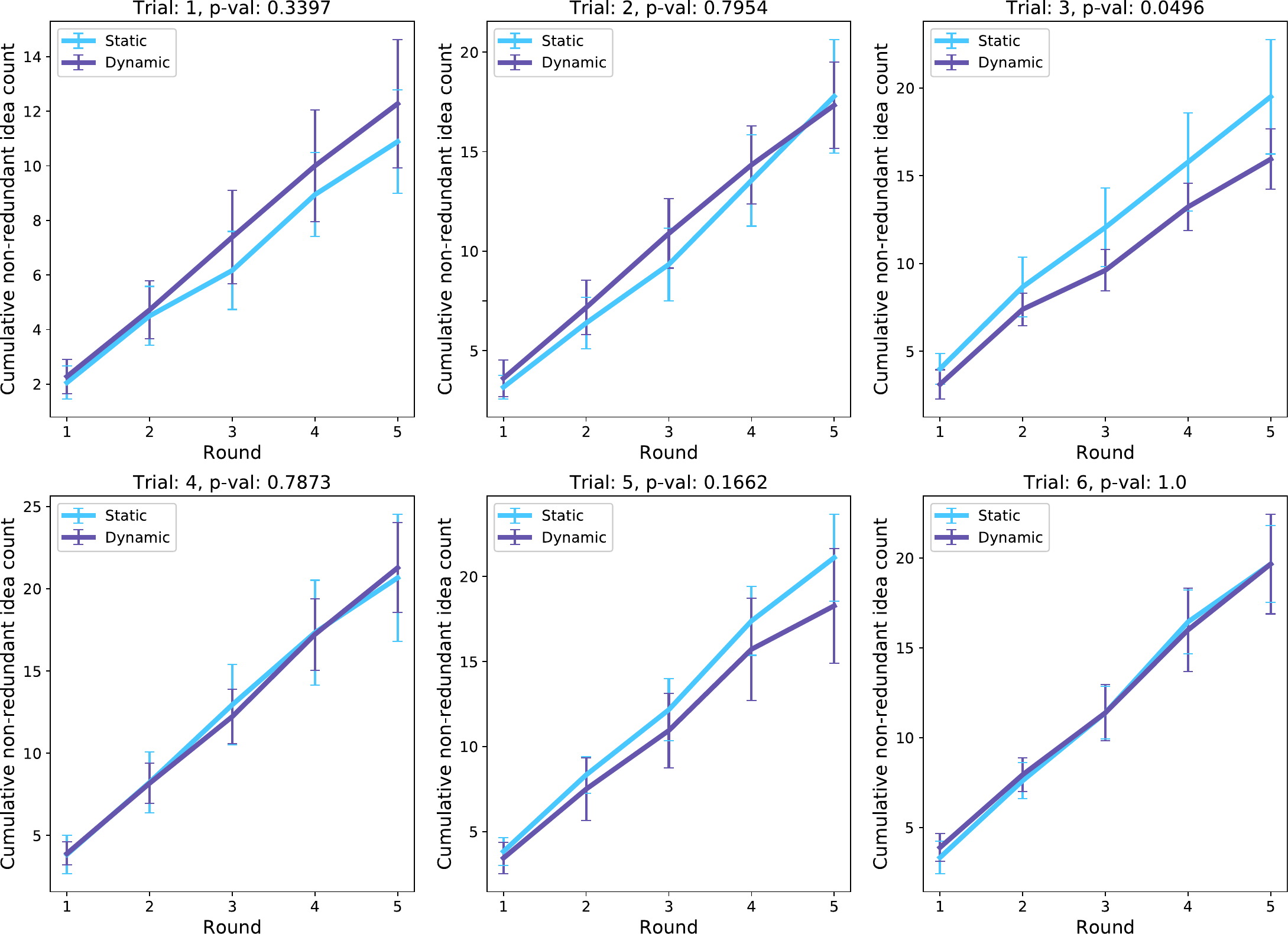}
    \caption[Trial-wise comparisons of non-redundant idea counts between static and dynamic egos]{Trial-wise comparisons of non-redundant idea counts between static and dynamic egos. 2-tailed tests are performed between the cumulative counts of static (s) and dynamic (d) conditions at the end of all $5$ rounds, as detailed in the following: Trial 1: $m_s=10.89$, $m_d=12.28$, $t(34)=-0.968$, $p=0.3397$, $95\%$ C.I. for $m_s-m_d=[-4.221, 1.444]$; Trial 2: $m_s=17.78$, $m_d=17.33$, $t(34)=0.261$, $p=0.7954$, $95\%$ C.I. for $m_s-m_d=[-2.914, 3.803]$; Trial 3: $m_s=19.5$, $m_d=15.94$, $t(34)=2.036$, $p=0.0496$, $95\%$ C.I. for $m_s-m_d=[0.106, 7.005]$; Trial 4: $m_s=20.67$, $m_d=21.28$, $t(34)=-0.272$, $p=0.7873$, $95\%$ C.I. for $m_s-m_d=[-5.050, 3.828]$; Trial 5: $m_s=21.11$, $m_d=18.28$, $t(34)=1.415$, $p=0.1662$, $95\%$ C.I. for $m_s-m_d=[-1.122, 6.789]$; Trial 6: $m_s=19.67$, $m_d=19.67$, $t(34)=0.0$, $p=1.0$, $95\%$ C.I. for $m_s-m_d=[-3.280, 3.280]$. Whiskers represent $95\%$ C.I.}
    \label{SI5}
\end{figure}

\begin{figure}[H]
    \centering
    \includegraphics[width=1\linewidth]{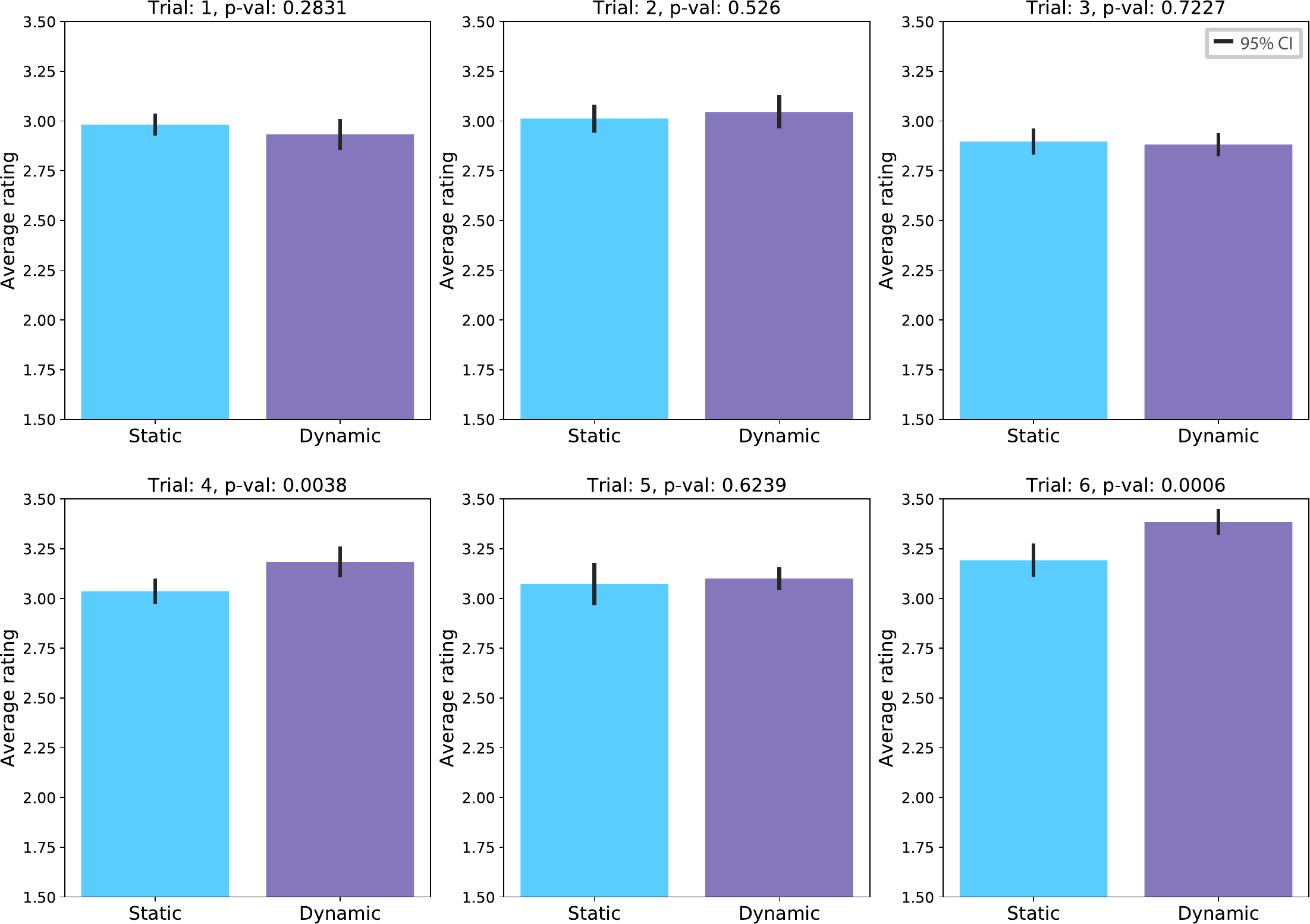}
    \caption[Trial-wise comparisons of average novelty ratings between dynamic and static egos]{Trial-wise comparisons of average novelty ratings between dynamic and static egos. 2-tailed tests are performed between the average novelty ratings of dynamic (d) and static (s) conditions over all $5$ rounds, as detailed in the following: Trial 1: $m_d=2.93$, $m_s=2.98$, $t(34)=-1.091$, $p=0.283$, $95\%$ C.I. for $m_d-m_S=[-0.137, 0.04]$; Trial 2: $m_d=3.05$, $m_s=3.01$, $t(34)=0.641$, $p=0.526$, $95\%$ C.I. for $m_d-m_s=[-0.069, 0.136]$; Trial 3: $m_d=2.88$, $m_s=2.9$, $t(34)=-0.358$, $p=0.723$, $95\%$ C.I. for $m_d-m_s=[-0.097, 0.067]$; Trial 4: $m_d=3.18$, $m_s=3.04$, $t(34)=3.107$, $p=0.0038$, $95\%$ C.I. for $m_d-m_s=[0.054, 0.241]$; Trial 5: $m_d=3.1$, $m_s=3.07$, $t(34)=0.495$, $p=0.624$, $95\%$ C.I. for $m_d-m_s=[-0.084, 0.14]$; Trial 6: $m_d=3.38$, $m_s=3.19$, $t(34)=3.801$, $p=0.00057$, $95\%$ C.I. for $m_d-m_s=[0.092, 0.292]$.}
    \label{SI6}
\end{figure}

\begin{figure}[H]
    \centering
    \includegraphics[width=1\linewidth]{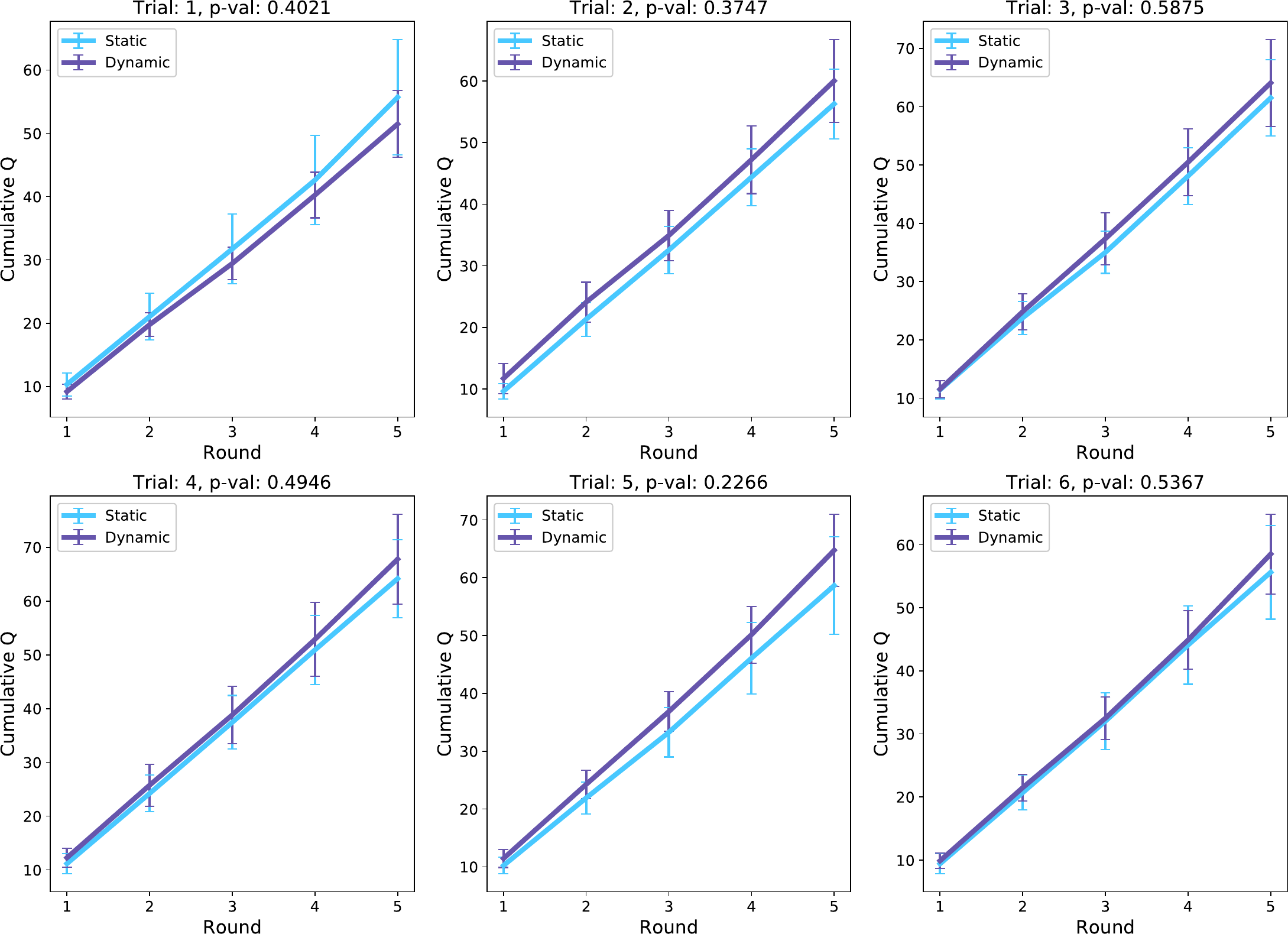}
    \caption[Trial-wise comparisons of creativity quotients between static and dynamic egos]{Trial-wise comparisons of creativity quotients between static and dynamic egos. 2-tailed tests results between the cumulative $Q$ counts of the static (s) and dynamic (d) conditions at the end of all $5$ rounds is given in the following: Trial 1: $m_s=55.71$, $m_d=51.47$, $t(34)=0.848$, $p=0.402$, $95\%$ C.I. for $m_s-m_d=[-5.628, 14.104]$; Trial 2: $m_s=56.28$, $m_d=60.03$, $t(34)=-0.9$, $p=0.375$, $95\%$ C.I. for $m_s-m_d=[-11.976, 4.481]$; Trial 3: $m_s=61.52$, $m_d=64.08$, $t(34)=-0.548$, $p=0.588$, $95\%$ C.I. for $m_s-m_d=[-11.833, 6.695]$; Trial 4: $m_s=64.17$, $m_d=67.8$, $t(34)=-0.69$, $p=0.495$, $95\%$ C.I. for $m_s-m_d=[-14.01, 6.752]$; Trial 5: $m_s=58.65$, $m_d=64.76$, $t(34)=-1.232$, $p=0.227$, $95\%$ C.I. for $m_s-m_d=[-15.912, 3.689]$; Trial 6: $m_s=55.64$, $m_d=58.53$, $t(34)=-0.624$, $p=0.537$, $95\%$ C.I. for $m_s-m_d=[-12.033, 6.254]$. Whiskers denote $95\%$ C.I.}
    \label{SI7}
\end{figure}

Figures~\ref{SI1} through \ref{SI7} present additional results from the analysis, as referred to from the main manuscript.

\newpage

\newpage
\section{Simulation model}
\begin{figure}
    \centering
    \includegraphics[width=0.9\linewidth]{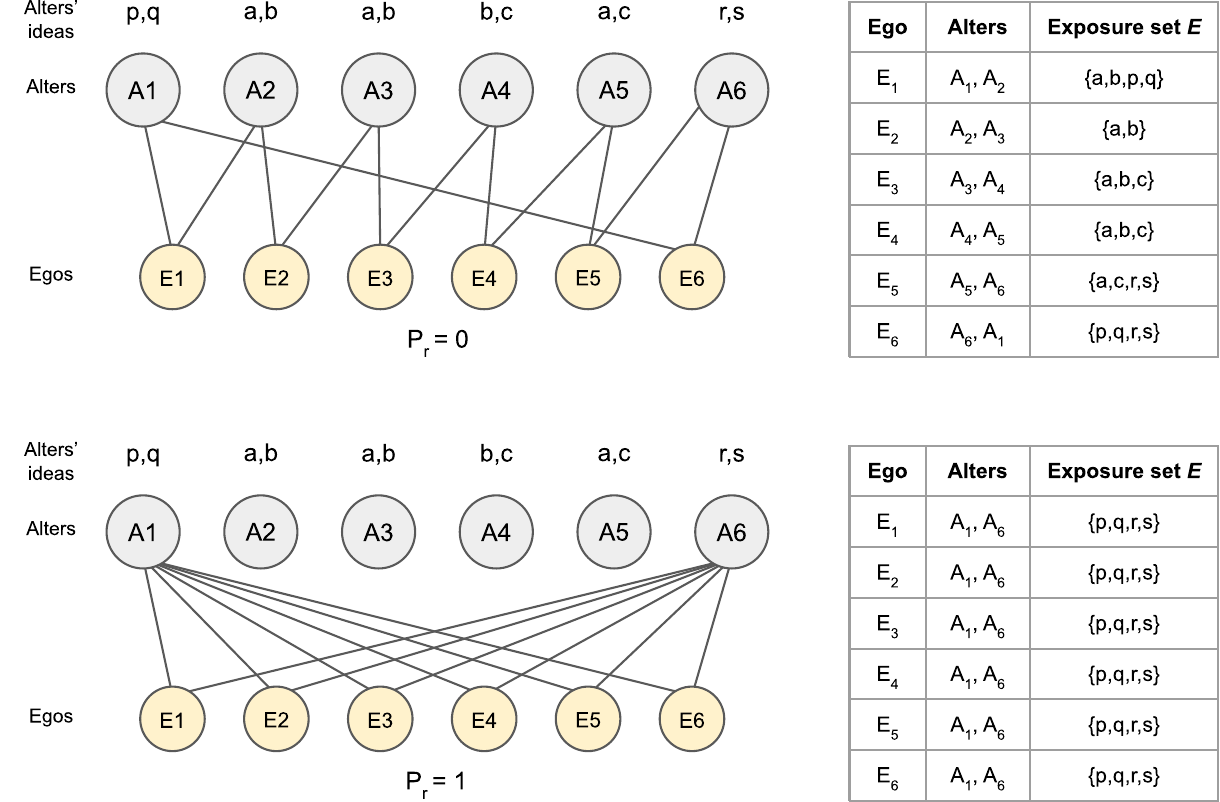}
    \caption[Simulation: Evolution of exposure set as the network rewires]{(\textit{Top row}) Simulation of the initial condition of the bipartite network (rewiring probability $P_r=0$). One realization of the stimuli idea set is shown here, where alters A1 and A6 generated non-redundant ideas (p, q and r, s respectively). Alters A2 through A5 generated ideas a, b and c, which are not unique and were submitted by multiple alters. Thus, A1 and A6 are the top-performing alters here. The egos are connected to the alters in the same pattern as used in the original experiment. $6$ egos are shown for demonstration purposes. The table to the left shows the computation of the exposure sets of the egos. (\textit{Bottom row}) The evolved network for $P_r=1$, where all the egos follow the same top-performing alters. This results in making all of the egos' exposure sets the same, as shown in the table on the left.}
    \label{SI12}
\end{figure}

We simulate the study outcomes using three key building blocks: (A) the network rewiring dynamics, (B) the cognitive stimulation mechanisms, and (C) the inter-ego similarities. In (A), we generate the initial network condition, the alters' ideas, the egos' exposure sets and the evolution of those exposure sets that stem from network rewiring. For generating the stimulated ideas based on these exposure sets, in (B), we abstract the cognitive mechanisms using linear, sub-linear and super-linear stimulation functions. Finally, in (C), we explore the outcomes in two corner cases of inter-ego similarities: full similarity and no similarity. We describe each block in detail below. We make our code available for easy replication of the model.

\subsection*{(A) Capturing the network dynamics}
\textbf{Network initialization.} Here, we adopt the same bipartite network settings as used in the empirical explorations. We first consider $m=6$ alters and $n=18$ egos, and initialize their connections in the same initial pattern as the original experiment. Each of the alters $i$ has an idea set $A_i$, which is used as the stimuli for the egos. Later, we experiment with larger networks that have $m=18$, $60$, and $600$, with $n=3m$ for each of those. We connect each ego to $2$ alters.

\textbf{Stimuli set generation.} Following the empirical observations in our study, we generate idea-sets $A_i$ for alters $i$ such that some of the alters have larger unique idea counts than others (popular and unpopular alters, respectively). To simulate this, we start with two pools (sets) of symbols representing unique ideas: $U_1$  and $U_2$. By having $|U_1|<<|U_2|$, we ensure that ideas sampled with replacement from $U_1$ will be more common than those from $U_2$. In other words, we simulate $U_1$ to include ideas that occur to people with a high probability, and $U_2$ to consist of rare ideas.

We assume that each alter $i$ generates a fixed number of $|A|$ ideas. Each idea in $A_i$ comes from pool $U_1$ with probability $\alpha_i$, or from $U_2$ with probability $1-\alpha_i$. For a random one-third of the alters, we take $0\leq \alpha_i \leq 0.5$ (high-performing alters), and for others $0.5<\alpha\leq1$ (low-performing alters). This makes the idea sets $A_i$ non-uniform, with the high-performing alters having a higher unique idea count than the low-performing alters, as shown in the top row of Figure~\ref{SI12}.

\textbf{Exposure set calculation.} For each ego $j$, we take the set of ideas they are exposed to as the exposure set $E_j = A_{i_1} \cup A_{i_2}$, where alters $i_1$ and $i_2$ are ego $j$'s peers.

\textbf{Evolution of exposure set.} With time (e.g., with rounds in our study), the egos in the dynamic condition can rewire their connections to the alters, which the static egos cannot. In the empirical results, we saw that the connection changes per ego dropped with time ($p<1e-4$ for the negative slope) as more egos followed the high-performing popular alters. We define a rewiring probability $P_r$ that captures how much the network deviates from its initial configuration ($P_r=0$) to the extreme case where two popular alters win the attention of all the egos ($P_r=1$). Therefore, instead of simulating the dynamic network through time to explore its temporal effects, we can equivalently sweep over the rewiring probability $P_r$ and explore its effects on the exposure sets of the egos. Figure~\ref{SI12} shows the idea. With higher $P_r$, the exposure sets of the egos become more uniform, as even the rare stimuli ideas from pool $U_2$ become common due to increased exposure.

\subsection*{(B) Capturing cognitive stimulation}

Given the exposure set $E_j$, an ego $j$ can generate the following: with probability $p_1$, s/he can generate ideas that are substantially inspired/stimulated by ideas from the exposure set, with probability $p_2$ s/he can generate ideas with negligible or no stimulation from the exposure set ideas, and with probability $p_3$ s/he can generate ideas that are inspired by the exposure set but do not fulfill the study requirements of being substantially different than the stimuli and also feasible. For our purposes of exploring the effects of the network dynamics, we can set $p_2=p_3=0$, which makes $p_1=1$. In other words, we are assuming that an ego only generates ideas that are inspired by the exposure set. Any effect from $p_2$ and $p_3$ should occur similarly in both static and dynamic conditions as the participants are randomly placed, and therefore act as mere random noise that we set to $0$. This leads to the set of stimulated ideas for ego $j$, $S_j = \{e'_1\}\cup\{e'_2\}\cup...\cup\{e'_k\} $ where each idea in the exposure set $e_k \in E_j$ leads to a set of ideas $S^{(e_k)}=\{e'_k\}$, and the union of all such idea sets from all $e_k\in E_j$ are contained in $S_j$.

The empirical results show a positive stimulation of ideas in the dynamic and static conditions compared to the solo condition (no stimuli). Therefore we can reasonably ignore the possibility that a stimulus can hurt the ideation process (negative association between $|E|$ and $|S|$). Also, our choice of having $p_1=1$ in the previous paragraph gets rid of the possibility of no association between $|E|$ and $|S|$. This leaves a positive stimulation effect, captured by a positive association between $|E|$ and $|S|$. 

As argued in the main manuscript, less overlap between an ego's own ideas and his/her alters' ideas can help in stimulating further novel ideas in the ego. Again, the rarer a stimulus idea $e$ is, the less overlap can be expected to exist between $e$ and the ego's own ideas, which can lead to a higher chance of stimulation. We measure the rarity of each stimulus idea as $R_{e} = 1- \frac{\textrm{Number of times the idea was submitted by the alters}}{\textrm{total number of alters' ideas}}$. Therefore, we have the number of ideas stimulated by $e$, $|S^{(e)}| \propto f(R_{e})$, where $f$ is a stimulation function. We consider three cases of this stimulation relation: (1) linear: $|S^{(e)}| = kR_{e}$, (2) sub-linear: $|S^{(e)}| = k\sqrt{R_{e}}$, (3) super-linear: $|S^{(e)}| = kR_{e}^2$, where $k$ is a proportionality constant.

\subsection*{(C) Capturing inter-ego similarity}
Every ego $j$ generates stimulated ideas $S_j$ independently of other egos. However, when the network evolves such that the high-performing alters become highly popular (high rewiring probability $P_r$), the exposure sets of the egos can become similar. We consider two extreme cases in this regard: (1) No similarity: every ego $j$ with the same stimulus idea $e$ generates completely different stimulated ideas in $S^{(e)}$, and (2) Full similarity: every ego $j$ with the same stimulus idea $e$ generates exactly the same stimulated ideas in $S^{(e)}$.

\begin{figure}
    \centering
    \includegraphics[width=0.6\linewidth]{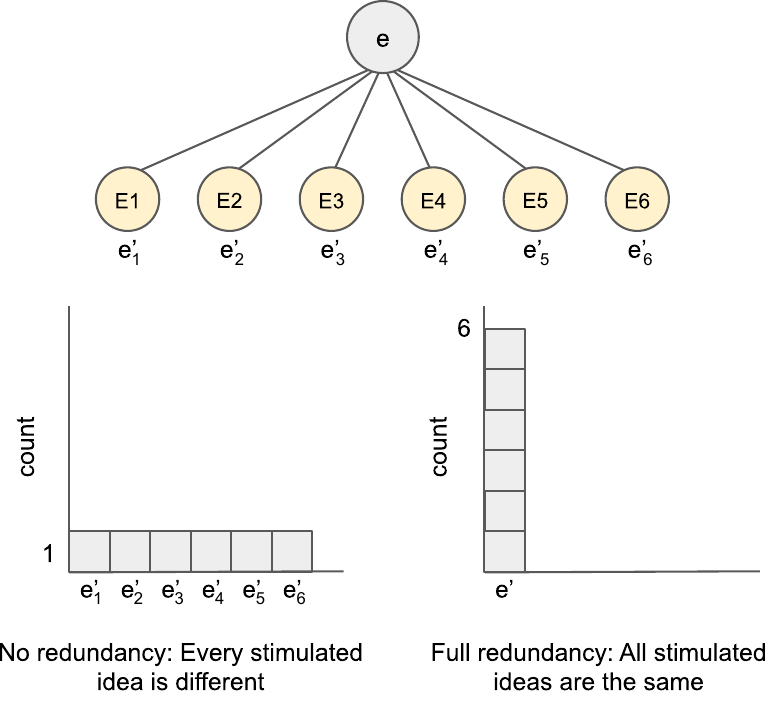}
    \caption[Simulation: Intuition behind inter-ego similarity]{(\textit{Top row}) An illustration of one stimulus $e$ being shown to $6$ independent egos, where the egos generate one stimulated idea each.  (\textit{Bottom row}) Two extreme cases: (1) No similarity/redundancy, where each stimulated idea is unique, and (2) Full similarity/redundancy, where all the stimulated ideas turn out to be the same. The dynamic network suffers in case of increased similarity, since the rewiring process exposes an increased number of people to the same stimulus $e$.}
    \label{SI13}
\end{figure}

The first case will have the least network effect due to the complete uniqueness of every stimulated idea. But in the second case, the dynamic network will suffer from generating more redundant ideas among the participants. An example is shown in Figure~\ref{SI13}.

To evaluate performance of the alters, we set a non-redundancy threshold of $1$, as explained in the main manuscript. In other words, any idea that is generated by at most one alter is considered non-redundant. For the egos, we take an idea to be non-redundant if it is generated by at most $15\%$ of the number of egos on the simulation.

\subsection*{Experiments} 
We explore the following cases in our simulation:

\textbf{Network size.} We experiment with bipartite networks consisting of $m=\{6,18,60,600\}$ alters and $n=3m$ egos. Each ego is connected to $2$ alters using the same initial configuration pattern as the original experiment. The results of the four cases are shown in Figures~\ref{SI_sim6}-\ref{SI_sim600} respectively.

\textbf{Inter-ego similarity.} For each network size, we consider both of the cases of no inter-ego similarity and full similarity. In each of the Figures~\ref{SI_sim6}-\ref{SI_sim600}, the left column and right columns respectively show the two cases.

\textbf{Cognitive stimulation functions.} For each of the inter-ego similarity cases, we experiment with three stimulation functions, relating the rarity of the stimulus ideas to the number of novel ideas in sub-linear, linear and super-linear ways. These results are shown in the top, middle and bottom rows of the Figures~\ref{SI_sim6}-\ref{SI_sim600}.

\textbf{Rewiring probabilities.} For the dynamic condition, we sweep through the rewiring probability $P_r$ from $0$ (initial condition) to $1$ (all of the egos follow only the two most popular alters). For the static control, we keep $P_r$ fixed at $0$.

\subsection*{Results and Discussion}

\begin{figure}
    \centering
    \includegraphics[width=1\linewidth]{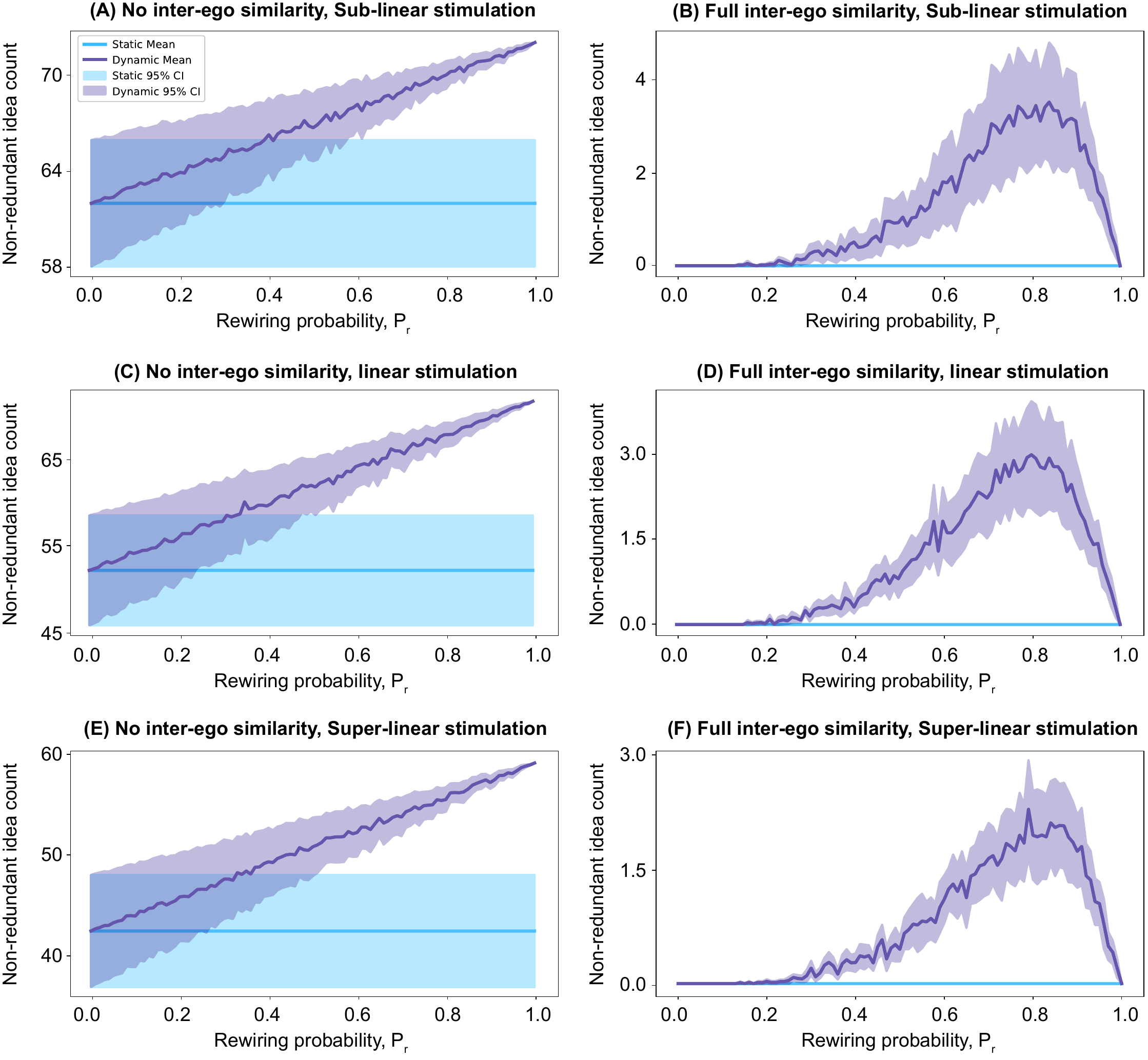}
    \caption[Simulation: Results for $m=6$ alters and $n=18$ egos]{Simulation results for $m=6$ alters and $n=18$ egos, computed for each of the three stimulation functions. The x-axis denotes rewiring probability $P_r$, where $P_r=0$ denotes the initial network structure and $P_r=1$ denotes the extreme case where all the egos follow the same two popular alters. The left column panels (A, C and E) show the simulation results for the case of no redundancy among the ideas generated by different egos in response to the same stimulus. The right column panels (B, D and F) show results for full redundancy cases. The top row, middle row and bottom row are the simulation results for the sub-linear, linear and super-linear stimulation functions, respectively. As can be seen, when there is no redundancy, the dynamic networks outperform the static ones as $P_r$ increases. However, when there is redundancy, the dynamic network suffers as more egos follow the same alters at higher $P_r$, eventually making all the stimulated ideas redundant and therefore not creative. Slope parameter $k=20$ has been used in the stimulation functions.}    
    \label{SI_sim6}
\end{figure}

\begin{figure}
    \centering
    \includegraphics[width=1\linewidth]{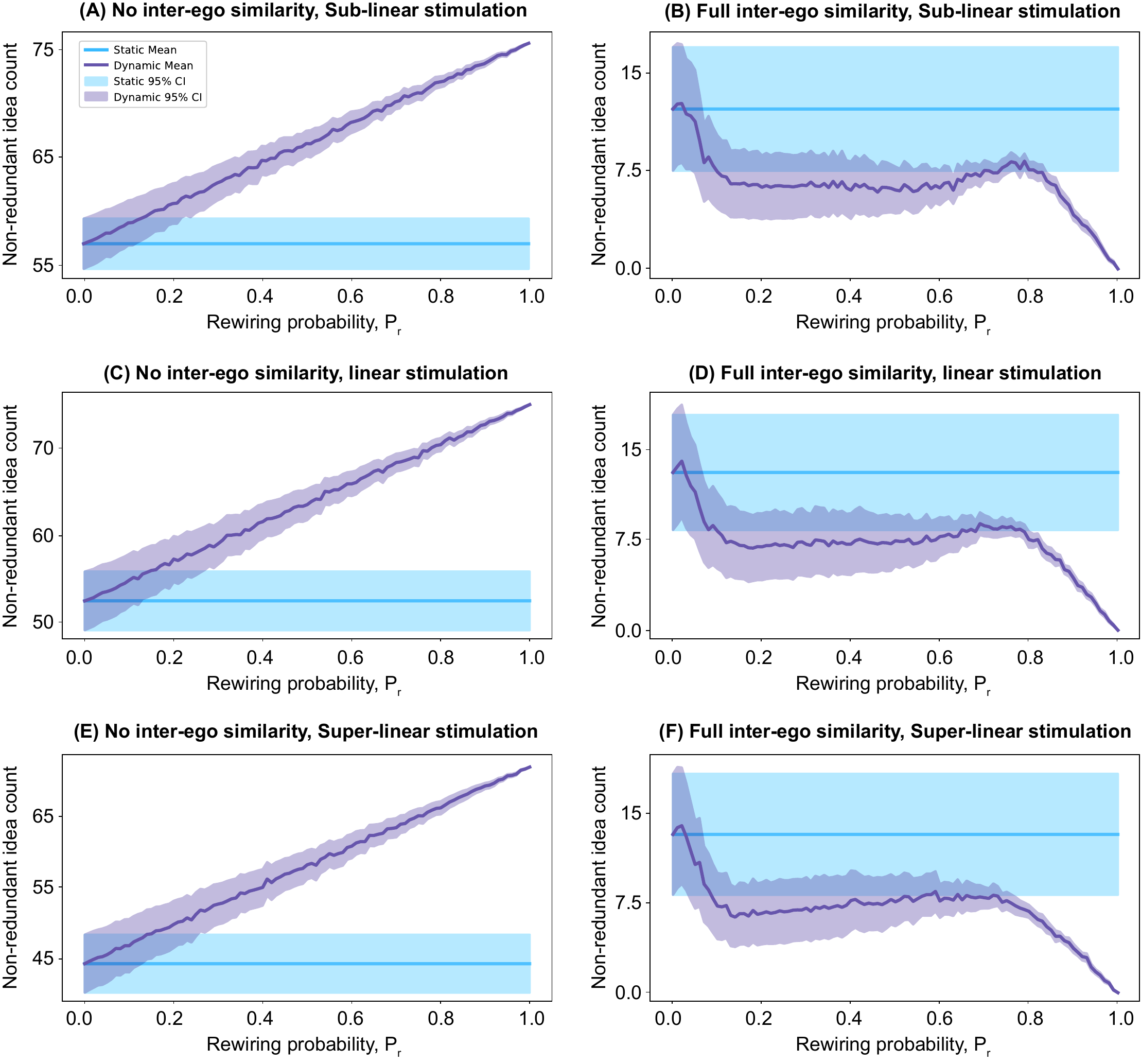}
    \caption[Simulation: Results for $m=18$ alters and $n=54$ egos]{Simulation results for $m=18$ alters and $n=54$ egos. }    
    \label{SI_sim18}
\end{figure}

\begin{figure}
    \centering
    \includegraphics[width=1\linewidth]{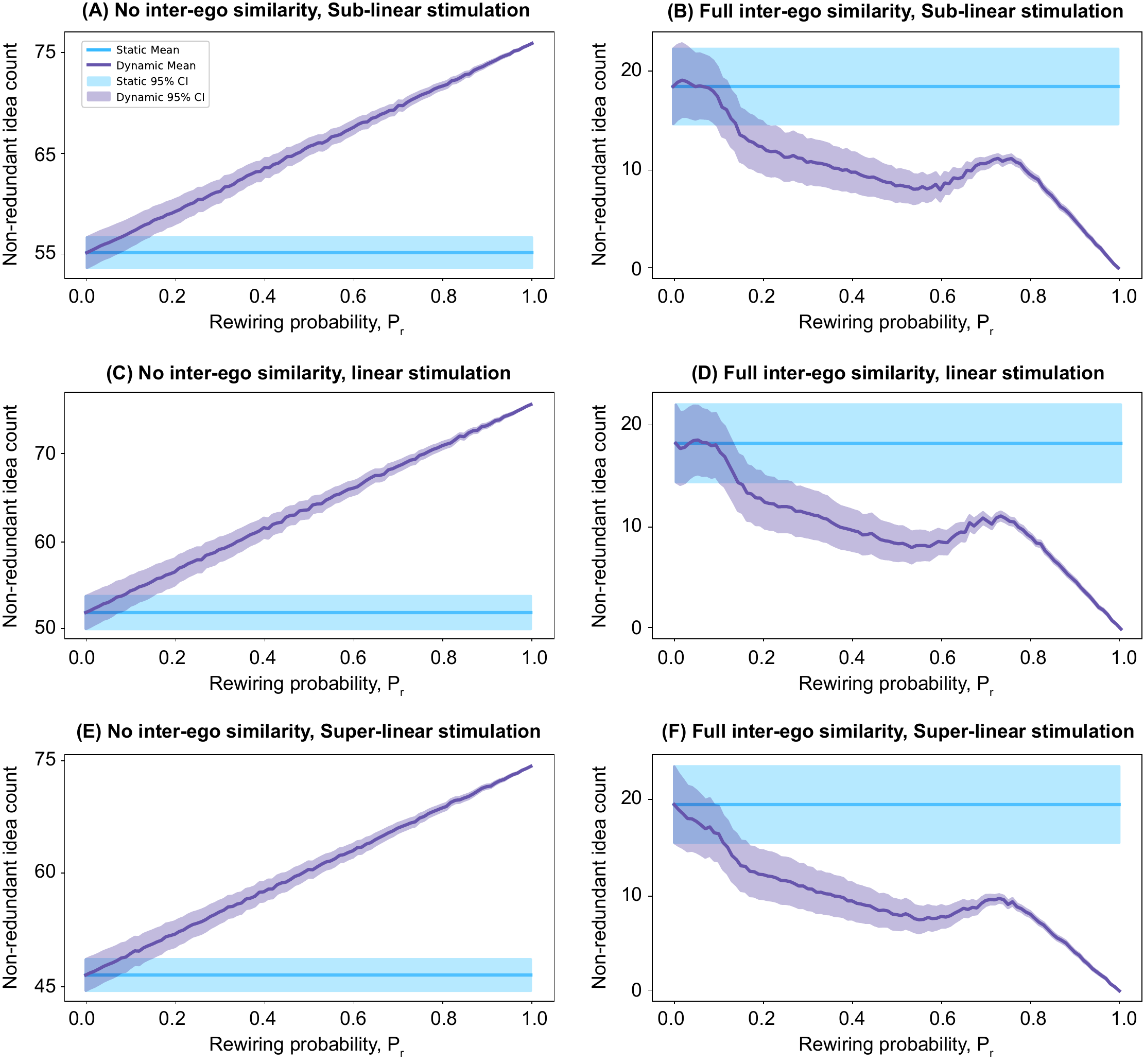}
    \caption[Simulation: Results for $m=60$ alters and $n=180$ egos]{Simulation results for $m=60$ alters and $n=180$ egos. }    
    \label{SI_sim60}
\end{figure}

\begin{figure}
    \centering
    \includegraphics[width=1\linewidth]{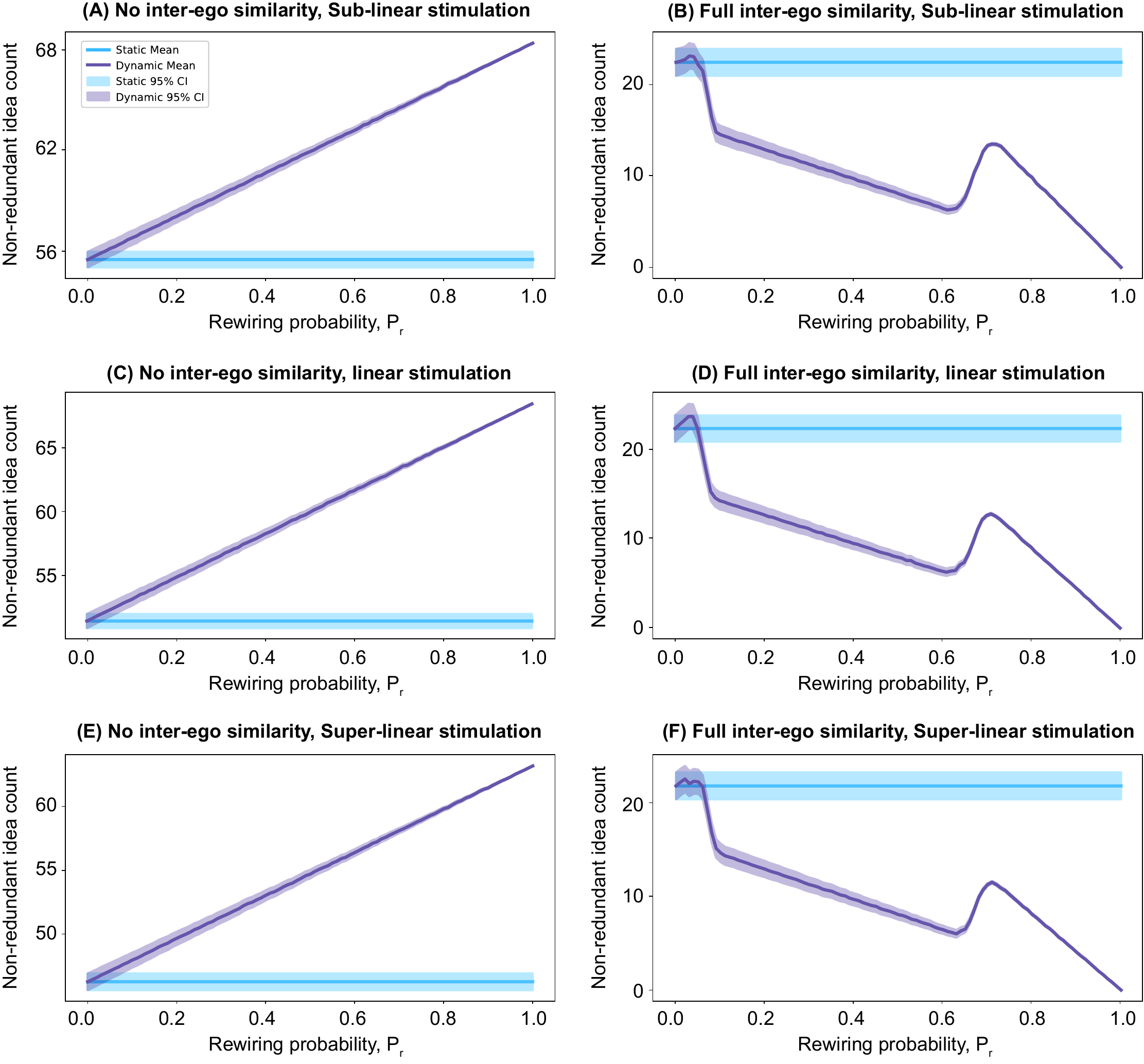}
    \caption[Simulation: Results for $m=600$ alters and $n=1800$ egos]{Simulation results for $m=600$ alters and $n=1800$ egos. }    
    \label{SI_sim600}
\end{figure}

The results are shown in Figures~\ref{SI_sim6}-\ref{SI_sim600}. When there is no similarity/redundancy among the egos' ideas generated in response to the same stimuli, the dynamic condition enjoys an advantage over the static condition as the rewiring probability $P_r$ increases. The network's performance maximizes at $P_r=1$. But when there is full redundancy, none of the ideas in the dynamic condition remains unique anymore as $P_r$ approaches $1$, thereby hurting the creative outcomes. This result is robust to various stimulation functions we chose, and also to network size.

The simulation highlights the roles played by the network dynamics and the cognitive stimulation mechanism in the creative ideation process. First, the rewiring process makes the stimuli set similar with time for the egos in the dynamic condition, which is a purely network-driven process. Second, the redundancy among the egos' ideas in response to the same stimulus also becomes a manifestation of the network dynamics, as the redundancy is initiated/facilitated by the egos' similar choices of peers. These two factors, taken together, negatively impact the creative outcomes in the dynamic condition. On the other hand, the stimulation process of the egos' ideas is driven by cognitive mechanisms. The various stimulation functions we experimented with ($f$) benefit the creative outcomes in varying degrees. However, as the simulation demonstrates, sufficient redundancy in the egos' ideas has the ability to overpower the cognitive stimulation benefits. In our empirical data, we find evidence of both of the network and cognitive factors to be present concurrently, which are captured by this simulation model.

\newpage
\section{Study interface}
The study was conducted with approval from the University IRB. No personally identifiable information was collected from the participants. The web interfaces used in the experiment are shown below, using pseudo usernames. Some of the materials are redacted to ensure copyright compliance of using materials from Guilford's Alternate Uses test.

\begin{figure}[H]
    \centering
    \includegraphics[width=1\linewidth]{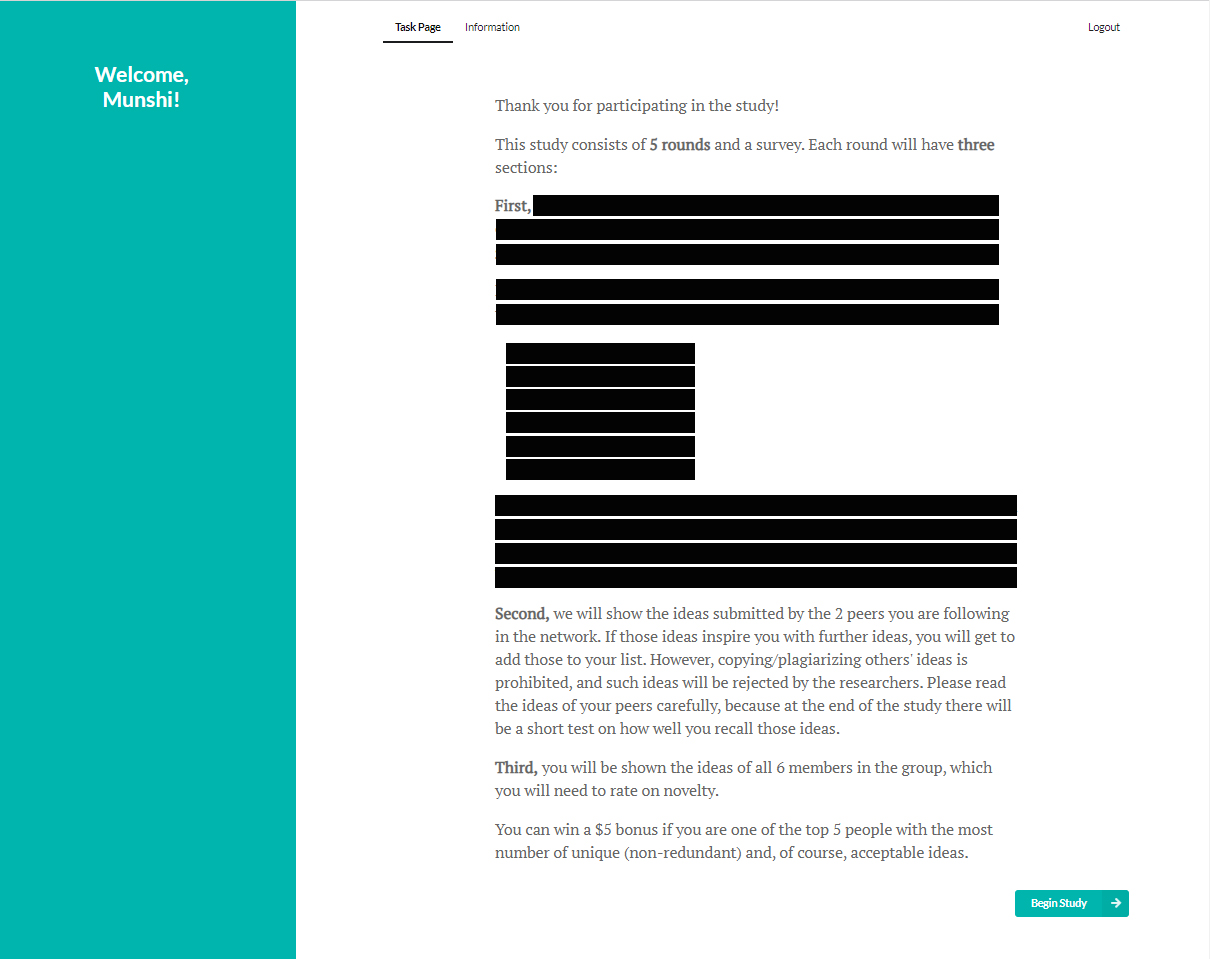}
    \caption[Study interface: Instruction page for static egos]{Instruction page for the egos of the static condition. Here, the first point is redacted to ensure copyright compliance of using the Guilford's test. This first point provides instructions for idea generation with examples. For the alters and solo participants, only the first point was shown.}    
    \label{SI001}
\end{figure}

\begin{figure}
    \centering
    \includegraphics[width=1\linewidth]{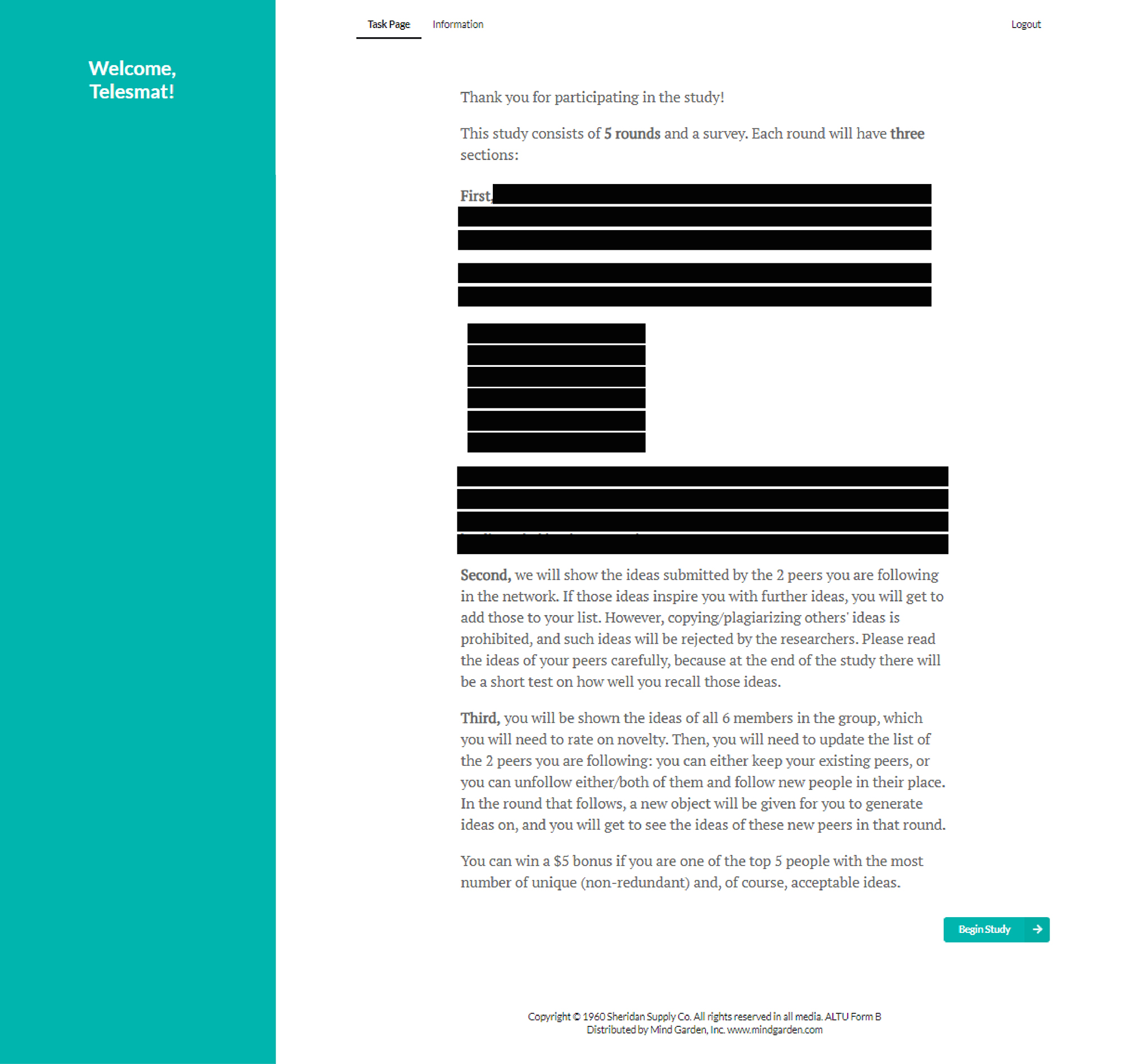}
    \caption[Study interface: Instruction page for dynamic egos]{Instruction page for the egos of the dynamic condition.}    
    \label{SI002}
\end{figure}

\begin{figure}
    \centering
    \includegraphics[width=1\linewidth]{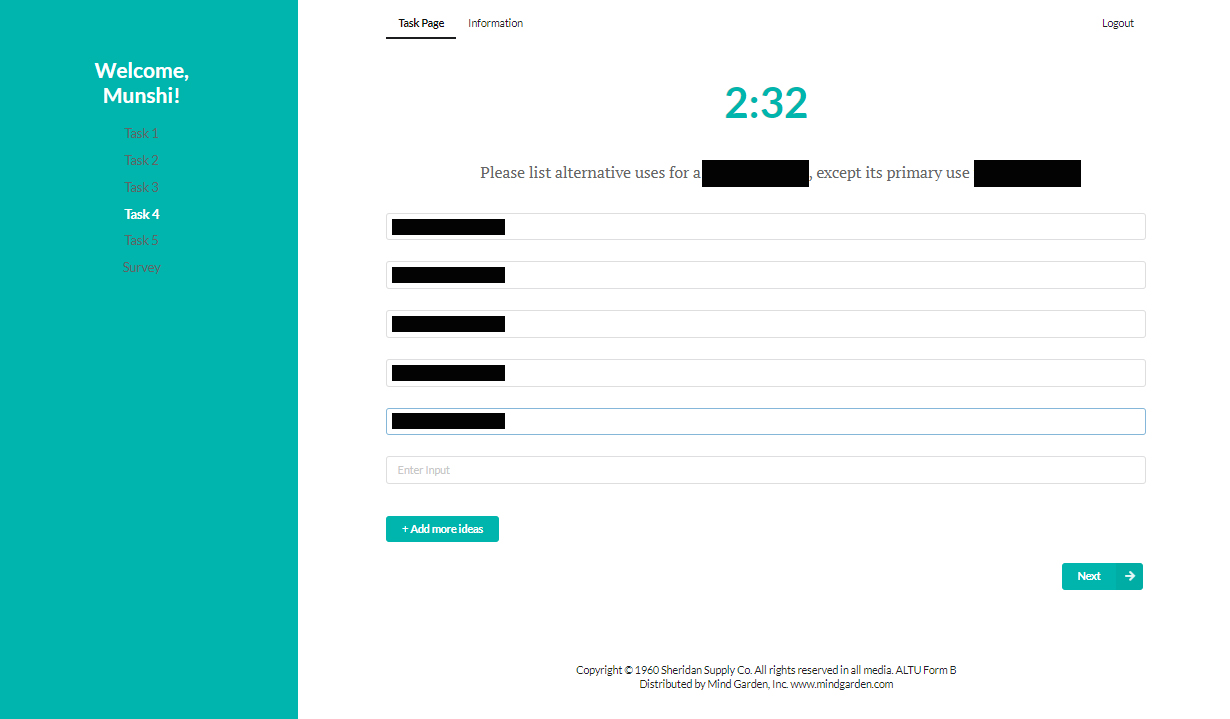}
    \caption[Study interface: Initial idea submission interface]{Initial idea submission interface. This was used in turn-1 for the egos of static and dynamic conditions, as well as for the alters and solo participants.}    
    \label{SI003}
\end{figure}

\begin{figure}
    \centering
    \includegraphics[width=1\linewidth]{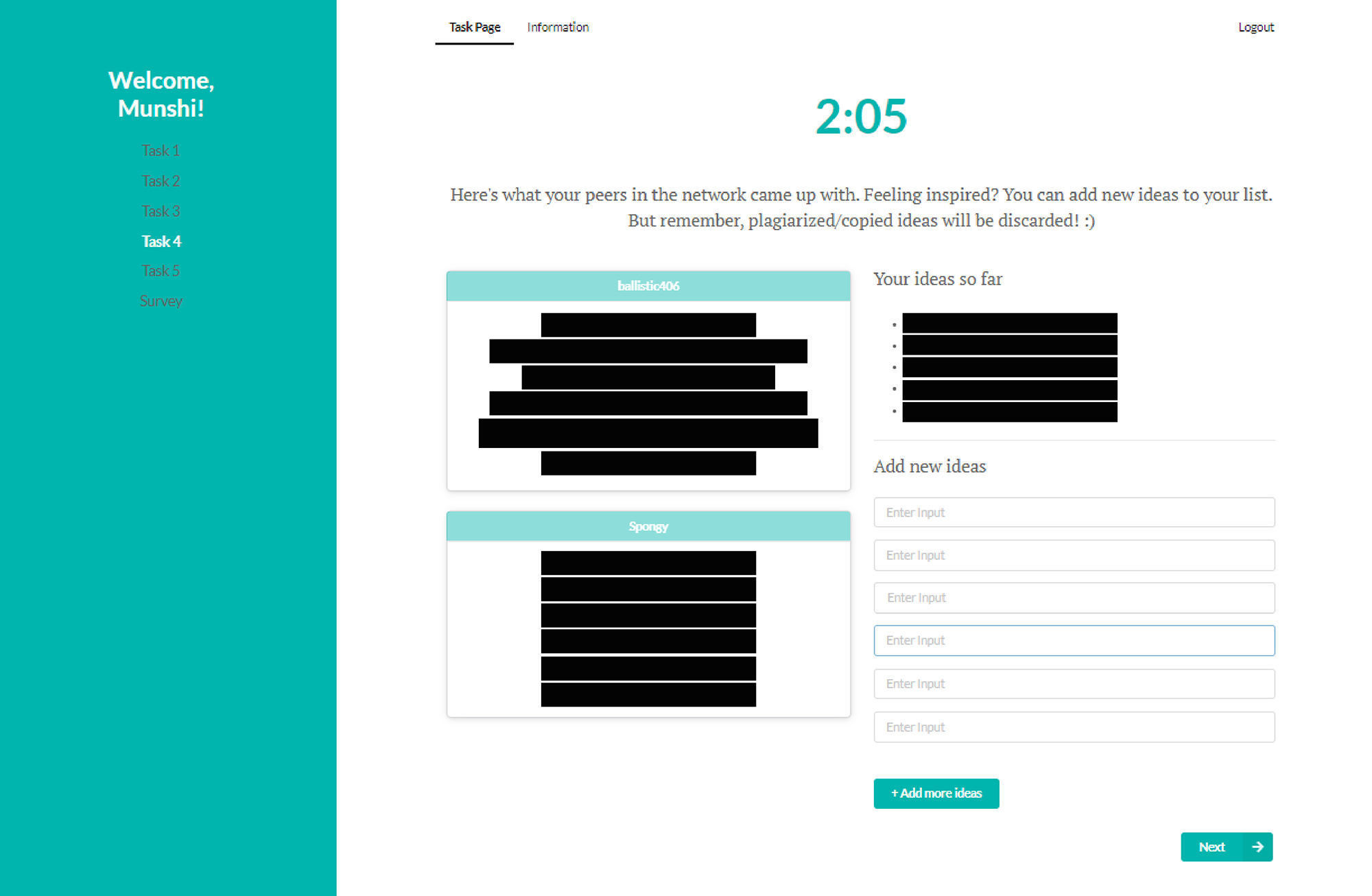}
    \caption[Study interface: Turn-2 interface for the egos of static and dynamic conditions.]{Turn-2 interface for the egos of static and dynamic conditions. The alters' ideas are shown on the left-side cards.}    
    \label{SI004}
\end{figure}

\begin{figure}
    \centering
    \includegraphics[width=1\linewidth]{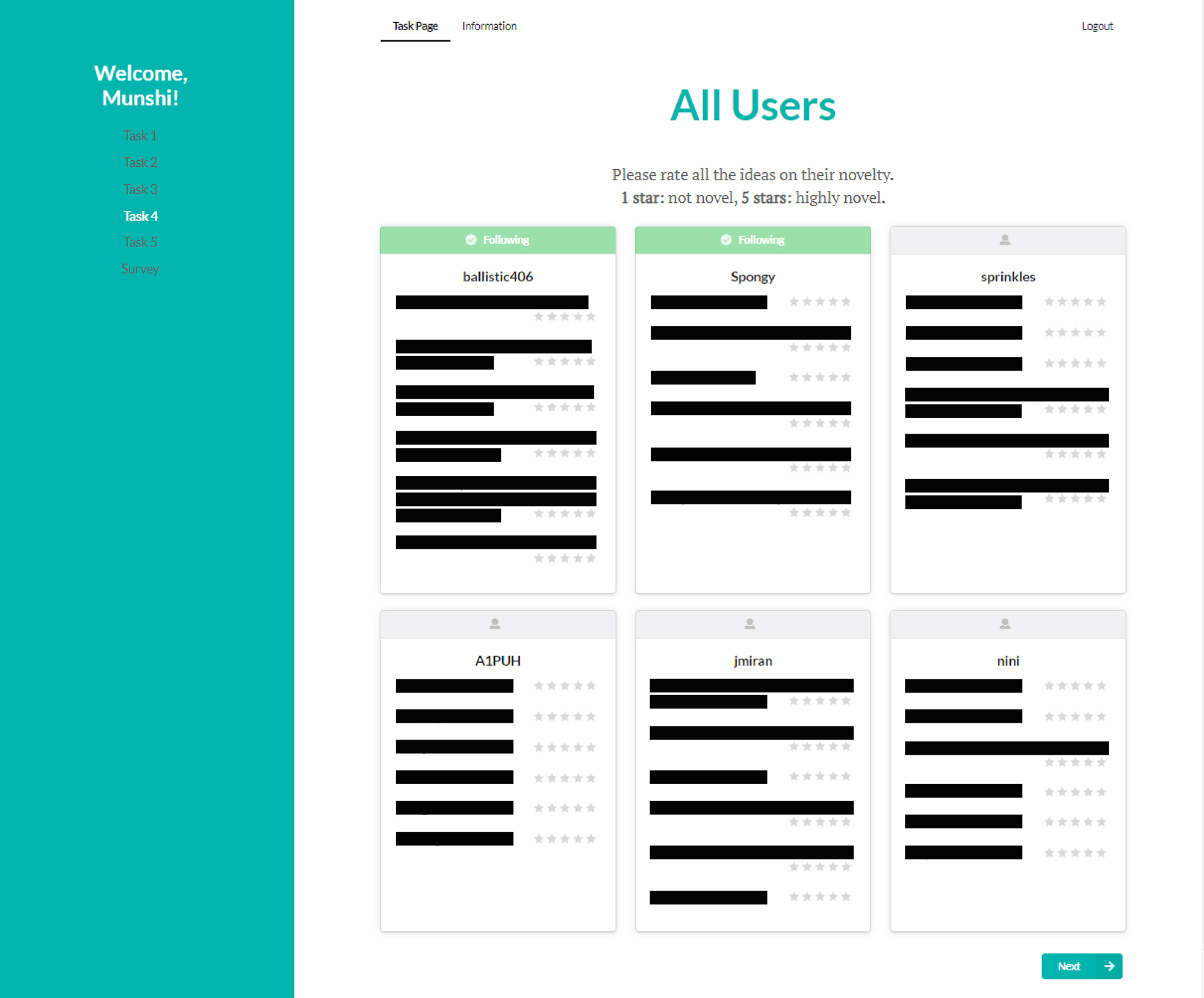}
    \caption[Study interface: Rating interface for static egos]{Rating interface for the egos in the static condition. The egos rated the ideas of all $6$ alters in the respective trial.}    
    \label{SI005}
\end{figure}

\begin{figure}
    \centering
    \includegraphics[width=1\linewidth]{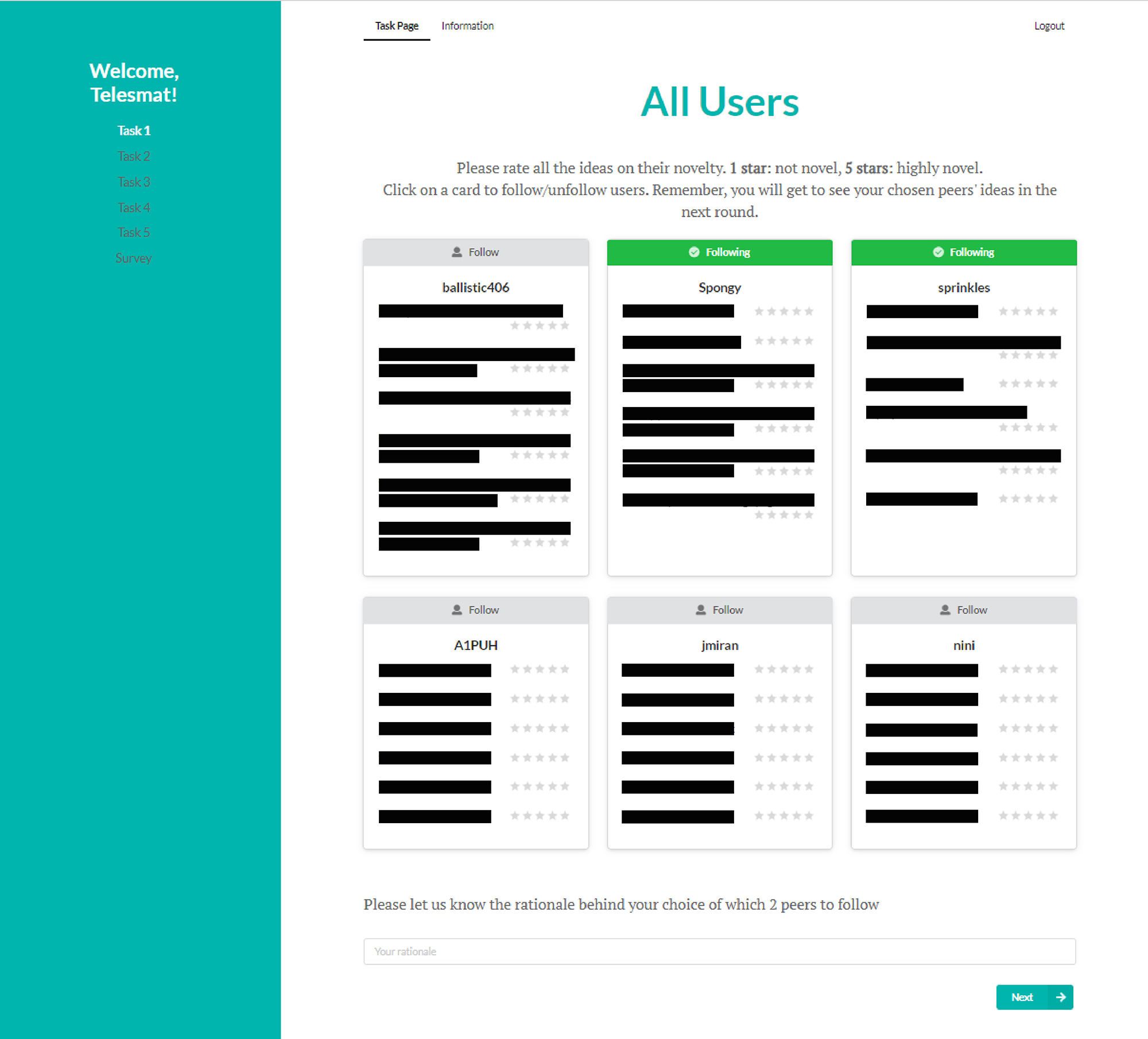}
    \caption[Study interface: Rating and rewiring interface for dynamic egos]{Rating and following/unfollowing interface for the egos in the dynamic condition.}    
    \label{SI006}
\end{figure}

\newpage

\section{Supplementary tables}

\begin{table}[H]
    \centering
    \caption[Performance comparisons between popular ($p$) and unpopular ($u$) alters]{Performance comparisons between popular ($p$) and unpopular ($u$) alters. 2-tailed tests. Data aggregated over all trials, $n_p =13$, $n_u=23$.}
    \begin{tabular}{|c|c|c|c|c|c|c|}
    \hline
    \textbf{Metric} & \textbf{$m_p$} & \textbf{$m_u$} & \textbf{$t$-statistic} & \textbf{df} & \textbf{$p$} & \textbf{$95\%$ C.I. for $m_p-m_u$} \\
    \hline
Non-redundant Idea Counts & $23.8$ & $14.4$ & $7.291$ & $34$ & $<0.001$ & $[6.9,12]$ \\
\hline
Average Ratings & $3.1$ & $2.6$ & $5.7$ & $34$ & $<0.001$ & $[0.3, 0.6]$ \\
\hline
Creativity Quotient & $57.8$ & $36.7$ & $5.81$ &$34$ & $<0.001$ & $[13.9, 28.2]$ \\
\hline
    \end{tabular}
    \label{pop_vs_unpop_performance}
\end{table}

\begin{table}[h]
\centering
\caption[Omnibus test results for the overlaps between the egos' turn-1 ideas and their alters' ideas]{Omnibus test results for analyzing the overlaps between the egos' turn-1 ideas and their alters' ideas. The overlap (Jaccard index) is the response variable. The analysis of variance of aligned rank transformed data is run on a mixed effects model with two factors: the number of popular alters of the egos (`Group factor', $3$ levels) and RoundID (`Round factor', $5$ levels). The degrees of freedom are specified using the Kenward-Roger method. Each RoundID has $n_r=216$ entries, one from each ego. Groups (i), (ii) and (iii), as defined in the main text, have $n_i=273$, $n_{ii}=476$, and $n_{iii}=331$ entries respectively.}
\begin{tabular}{|c|c|c|c|c|}
  \hline
  & Df & Df.res & $F$ & Pr($>F$) \\ 
  \hline
NumPopularAlters (Group factor) & $2$ & $669.84$ &  $66.526$ & $<2.22e-16$ \\ 
  \hline
  RoundID (Round factor) & $4$ & $866.07$ & $57.307$   & $<2.22e-16$ \\ 
    \hline
  NumPopularAlters:RoundID & $8$ & $940.38$ & $8.474$   & $3.561e-11$  \\ 
   \hline
\end{tabular}
\label{jaccart_omnibus}
\end{table}

\begin{table}[h]
\centering
\caption[Post-hoc analysis on the overlaps between the egos' turn-1 ideas and their alters' ideas]{Post-hoc analysis among the three levels in the Group factor from the fitted model reported in Table~\ref{jaccart_omnibus}. The degrees of freedom are specified using the Kenward-Roger method. The $p$-values are adjusted using Holm's sequential Bonferroni procedure.}
\begin{tabular}{|c|c|c|c|c|c|}
  \hline
 Contrast & Estimate & SE & df & $t$ &  $p$ \\ 
  \hline
Group (iii)-Group (ii) & $130$ & $23.4$ &  $574$ & $5.555$ & $<0.001$ \\ 
  \hline
  Group (iii)-Group (i) & $308$ & $26.8$ & $597$   & $11.481$ & $<0.001$ \\ 
    \hline
  Group (ii)-Group (i) & $178$ & $23.6$ & $887$   & $7.519$ & $<0.001$ \\ 
   \hline
\end{tabular}
\label{jaccart_posthoc}
\end{table}

\begin{table}[H]
    \centering
    \caption[Omnibus test results for analyzing the egos' turn-2 performances]{Omnibus test results for analyzing the egos' turn-2 performances. Three separate models are fitted for the three creativity metrics as the response variables. The analysis of variance of aligned rank transformed data is run on a mixed effects model with two factors: the number of popular alters of the egos (`Group factor', $3$ levels) and RoundID (`Round factor', $5$ levels). The degrees of freedom are specified using the Kenward-Roger method. Each RoundID has $n_r=216$ entries, one from each ego. Groups (i), (ii) and (iii), as defined in the main text, have $n_i=273$, $n_{ii}=476$, and $n_{iii}=331$ entries respectively.}
\begin{tabular}{|c|c|c|c|c|}
  \hline
  \multicolumn{5}{|c|}{\textbf{Metric: Non-redundant Idea Counts}}\\
  \hline
  & Df & Df.res & $F$ & Pr($>F$) \\ 
  \hline
NumPopularAlters (Group factor) & $2$ & $825.86$ &  $3.701$ & $0.025$ * \\ 
  \hline
  RoundID (Round factor) & $4$ & $862.74$ & $3.265$   & $0.011$ * \\ 
    \hline
  NumPopularAlters:RoundID & $8$ & $913.51$ & $0.513$   & $0.847$  \\ 
   \hline
   \hline
   \multicolumn{5}{|c|}{\textbf{Metric: Average Novelty Ratings}}\\
  \hline
  & Df & Df.res & $F$ & Pr($>F$) \\ 
  \hline
NumPopularAlters (Group factor) & $2$ & $535.47$ &  $11.852$ & $9.19e-6$ *** \\ 
  \hline
  RoundID (Round factor) & $4$ & $869.13$ & $8.361$   & $1.28e-6$ *** \\ 
    \hline
  NumPopularAlters:RoundID & $8$ & $973.46$ & $1.409$   & $0.189$  \\ 
   \hline
   \hline
   \multicolumn{5}{|c|}{\textbf{Metric: Creativity Quotient}}\\
  \hline
  & Df & Df.res & $F$ & Pr($>F$) \\ 
  \hline
NumPopularAlters (Group factor) & $2$ & $1036.36$ &  $6.657$ & $0.0013$ ** \\ 
  \hline
  RoundID (Round factor) & $4$ & $857.72$ & $14.836$   & $9.98e-12$ *** \\ 
    \hline
  NumPopularAlters:RoundID & $8$ & $880.30$ & $1.792$   & $0.075$  \\ 
   \hline
\end{tabular}
    \label{urq_omnibus}
\end{table}

\begin{table}[H]
\centering
\caption[Post-hoc analysis of the egos' turn-2 performances]{Post-hoc analysis among the three levels in the Group factor from the three fitted models reported in Table~\ref{urq_omnibus}. The degrees of freedom are specified using the Kenward-Roger method. The $p$-values are adjusted using Holm's sequential Bonferroni procedure.}
\begin{tabular}{|c|c|c|c|c|c|}
\hline
  \multicolumn{6}{|c|}{\textbf{Metric: Non-redundant Idea Counts}}\\
  \hline
 Contrast & Estimate & SE & df & $t$ &  $p$ \\ 
  \hline
Group (iii)-Group (ii) & $-70.0$ & $26.0$ &  $727$ & $-2.689$ & $0.022$ \\ 
  \hline
  Group (iii)-Group (i) & $-57.6$ & $29.7$ & $747$   & $-1.936$ & $0.106$ \\ 
    \hline
  Group (ii)-Group (i) & $12.5$ & $25.3$ & $1014$   & $0.495$ & $0.621$ \\ 
   \hline
   \hline
  \multicolumn{6}{|c|}{\textbf{Metric: Average Novelty Ratings}}\\
  \hline
 Contrast & Estimate & SE & df & $t$ &  $p$ \\ 
  \hline
Group (iii)-Group (ii) & $-79.4$ & $23.4$ &  $458$ & $-3.399$ & $0.0015$ \\ 
  \hline
  Group (iii)-Group (i) & $-127.8$ & $26.8$ & $483$   & $-4.759$ & $<0.0001$ \\ 
    \hline
  Group (ii)-Group (i) & $-48.3$ & $24.5$ & $721$   & $-1.971$ & $0.0491$ \\ 
   \hline
\hline
  \multicolumn{6}{|c|}{\textbf{Metric: Creativity Quotient}}\\
  \hline
 Contrast & Estimate & SE & df & $t$ &  $p$ \\ 
  \hline
Group (iii)-Group (ii) & $-85.814$ & $24.6$ &  $1016$ & $-3.495$ & $0.0015$ \\ 
  \hline
  Group (iii)-Group (i) & $-85.475$ & $27.9$ & $1022$   & $-3.062$ & $0.0045$ \\ 
    \hline
  Group (ii)-Group (i) & $0.339$ & $22.2$ & $1050$   & $0.015$ & $0.9878$ \\ 
   \hline
   \end{tabular}
\label{urq_posthoc}
\end{table}

\begin{table}[H]
    \centering
    \caption[Semantic dissimilarity comparisons among node-pairs]{Semantic dissimilarity comparisons among node-pairs with $0$, $1$ and $2$ common alter(s) at the end of the $5$\textsuperscript{th} round. Node-pairs with $2$ common alters were significantly less dissimilar than $0$ and $1$ common alter cases. $2$-tailed tests, data aggregated over all trials. Number of node-pairs: $n_2=170$, $n_1=464$, $n_0=284$, where the subscripts denote the number of common alters of the node pairs. All $p$-values are Bonferroni-corrected.}
    \begin{tabular}{|c|c|c|c|c|c|}
    \hline
     & \textbf{Means} & \textbf{$t$} & \textbf{df} & \textbf{$p$} & \textbf{$95\%$ C.I.} \\
    \hline
$2$ vs $0$ common alter(s) & $m_2=3.01$, $m_0=3.22$ & $-2.962$ & $452$ & $<0.01$ & $m_2-m_0=[-0.36, -0.07]$ \\
\hline
$2$ vs $1$ common alter(s) & $m_2=3.01$, $m_1=3.19$ & $-2.788$ & $632$ & $<0.02$ & $m_2-m_1=[-0.31, -0.05]$ \\
\hline
    \end{tabular}
    \label{echo_comb_table}
\end{table}

\begin{table}[H]
    \centering
    \caption[Individual-level comparisons of the total non-redundant idea counts of the egos]{Individual-level comparisons of the total non-redundant idea counts of the egos in the three study conditions. $2$-tailed tests, data aggregated over all trials. Number of observations: Dynamic: $n_d=108$, Static: $n_s=108$, Solo: $n_c=36$. All $p$-values are Bonferroni-corrected.}
    \begin{tabular}{|c|c|c|c|c|c|}
    \hline
     & \textbf{Means} & \textbf{$t$} & \textbf{df} & \textbf{$p$} & \textbf{$95\%$ C.I.} \\
    \hline
Dynamic (d) vs Solo (c) & $m_d=6.33$, $m_c=4.44$ & $2.7$ & $142$ & $<0.03$ &  $m_d-m_c=[0.52,3.26]$ \\
\hline
Static (s) vs Solo (c) & $m_s=6.77$, $m_c=4.44$ & $2.898$ & $142$ & $<0.02$ & $m_s-m_c=[0.75, 3.90]$ \\
\hline
    \end{tabular}
    \label{ind_table}
\end{table}

\newpage

\section{Power analysis for sample sizes}

\textbf{Link update patterns in the network evolution.} In Table~\ref{pow1}, we present the a priori power analysis of sample sizes using the t-test (difference between two independent means) given alpha, power and effect size. The effect sizes are determined from the observed data, while the allocation ratio is determined from the ratio of popular and unpopular alter counts observed in the data. To be conservative, we used $1.5$ times larger standard deviations within each group than the original data, to allow for a larger noise margin.

\begin{table}[H]
    \centering
    \caption[Power analysis: Link update patterns in the network evolution]{Power analysis: Link update patterns in the network evolution}
    \begin{tabular}{|c|c|c|c|c|c|c|c|}
    \hline
   \textbf{Metric}  & \textbf{Alpha} & \textbf{Power} & \textbf{Std} & \textbf{Allocation} & \textbf{Calculated} & \textbf{Calculated} & \textbf{Actual} \\
  & & & \textbf{factor} & \textbf{ratio ($N_p$/$N_u$)} & \textbf{effect size} & \textbf{sample size} & \textbf{sample size} \\
    \hline
Non-redun. Idea Ct. & $0.05$ & $0.8$ & $1.5$ & $0.565$ &  $1.72$ & $14$ & $36$ \\
\hline
Avg. Novelty Ratings & $0.05$ & $0.8$ & $1.5$ & $0.565$ & $1.48$ & $18$ & $36$ \\
\hline
Creativity Quotient & $0.05$ & $0.8$ & $1.5$ & $0.565$ & $1.33$ & $22$ & $36$ \\
\hline
    \end{tabular}
    \label{pow1}
\end{table}

\textbf{Exposure to high-performing alters is associated with better creative performances of the egos.} In Table~\ref{pow2}, we present the a priori power analysis of sample sizes using the F-test, given alpha, power and effect size. The effect sizes are determined from the observed data.

\begin{table}[H]
    \centering
    \caption[Power analysis: Exposure to high-performing alters is associated with better creative performances of the egos]{Power analysis: Exposure to high-performing alters is associated with better creative performances of the egos}
    \begin{tabular}{|c|c|c|c|c|c|}
    \hline
   \textbf{Metric}  & \textbf{Alpha} & \textbf{Power} &  \textbf{Calculated} & \textbf{Calculated} & \textbf{Actual} \\
  & & & \textbf{effect size} & \textbf{sample size} & \textbf{sample size} \\
    \hline
Jaccard Index             & $0.05$ & $0.8$ &  $0.32$ & $99$ & $1080$ \\
    \hline
Non-redundant Idea Counts & $0.05$ & $0.8$ &  $0.12$ & $714$ & $1080$ \\
\hline
Average Novelty Ratings   & $0.05$ & $0.8$  & $0.15$ & $408$ & $1080$ \\
\hline
Creativity Quotient       & $0.05$ & $0.8$  & $0.19$ & $261$ & $1080$ \\
\hline
    \end{tabular}
    \label{pow2}
\end{table}

\textbf{Following the same alters introduces semantic similarities in the egos’ ideas.} In Table~\ref{pow3}, we present the a priori power analysis of sample sizes using the F-test, given alpha, power and effect size. The effect sizes are determined from the observed data.

\begin{table}[H]
    \centering
    \caption[Power analysis: Following the same alters introduces semantic similarities in the egos’ ideas]{Power analysis: Following the same alters introduces semantic similarities in the egos’ ideas}
    \begin{tabular}{|c|c|c|c|c|}
    \hline
 \textbf{Alpha} & \textbf{Power} &  \textbf{Calculated} & \textbf{Calculated} & \textbf{Actual} \\
   & & \textbf{effect size} & \textbf{sample size} & \textbf{sample size} \\
    \hline
 $0.05$ & $0.8$ &  $0.104$ & $882$ & $918$ \\
    \hline
    \end{tabular}
    \label{pow3}
\end{table}

\textbf{Individual creative performance comparisons among various study conditions.} In Table~\ref{pow4}, we present the a priori power analysis of sample sizes using the F-test, given alpha, power and effect size. The effect sizes are determined from the observed data.

\begin{table}[H]
    \centering
    \caption[Power analysis: Individual creative performance comparisons among various study conditions]{Power analysis: Individual creative performance comparisons among various study conditions}
    \begin{tabular}{|c|c|c|c|c|}
    \hline
 \textbf{Alpha} & \textbf{Power} &  \textbf{Calculated} & \textbf{Calculated} & \textbf{Actual} \\
   & & \textbf{effect size} & \textbf{sample size} & \textbf{sample size} \\
    \hline
 $0.05$ & $0.8$ &  $0.2$ & $246$ & $252$ \\
    \hline
    \end{tabular}
    \label{pow4}
\end{table}

\end{document}